\documentclass[namedreferences]{solarphysics}
\usepackage{journalmacros}
\usepackage{apacite}
\usepackage[utf8x]{inputenc}
\usepackage{epsf}
\usepackage{longtable}
\usepackage{graphicx}
\usepackage{graphicx}
\usepackage{epstopdf}
\usepackage{color}
\usepackage{caption}
\usepackage{enumerate}
\usepackage{subfig}

\usepackage[hyperref,optionalrh,solaromanenum]{spr-sola-addons} 

\usepackage{graphicx}                    
\usepackage{color}                       
\usepackage{breakurl}                         

\begin{document}

\begin{article}

\begin{opening}

\title{Observations and Modelling of the Pre-Flare Period of the 29 March 2014 X1 Flare}

%
\author[addressref={1},corref,email={magnus.woods.15@ucl.ac.uk}]{\inits{M. M. }\fnm{M. M. }\lnm{Woods}\orcid{0000-0003-1593-4837}}
\author[addressref={1},corref,email={}]{\inits{L. K. }\fnm{L. K. }\lnm{Harra}\orcid{0000-0001-9457-6200}}
\author[addressref={1},corref,email={}]{\inits{S. A. }\fnm{S. A. }\lnm{Matthews}\orcid{0000-0001-9346-8179}}
\author[addressref={2},corref,email={}]{\inits{D. H. }\fnm{D. H. }\lnm{Mackay}\orcid{0000-0001-6065-8531}}
\author[addressref={1},corref,email={}]{\inits{S. }\fnm{S. }\lnm{Dacie}\orcid{ 0000-0001-7572-2903}}
\author[addressref={1},corref,email={}]{\inits{D. M. }\fnm{D. M. }\lnm{Long}\orcid{0000-0003-3137-0277}}
%
\runningauthor{M. M. Woods \textit{et al.}}
\runningtitle{Pre-flare Observations and Modelling}

\address[id={1}]{Mullard Space Science Laboratory, University College London, Holmbury St. Mary, Dorking, Surrey, RH5 6NT, UK}
\address[id={2}]{School of Mathematics and Statistics, University of St Andrews, North Haugh, St Andrews, Fife, FY16 9SS, UK}
\begin{abstract}
On the 29 March 2014 NOAA active region\,(AR) 12017 produced an X1 flare which was simultaneously observed by an unprecedented number of observatories. We have investigated the pre-flare period of this flare from 14:00\,UT until 19:00\,UT using joint observations made by the \textit{Interface Region Imaging Spectrometer} (\textrm{IRIS}) and the \textit{Hinode Extreme Ultraviolet Imaging Spectrometer} (\textrm{EIS}). Spectral lines providing coverage of the solar atmosphere from chromosphere to the corona were analysed to investigate pre-flare activity within the AR. The results of the investigation have revealed evidence of strongly blue-shifted plasma flows, with velocities up to 200\,km\,s$^{-1}$, being observed 40 minutes prior to flaring.  These flows are located along the filament present in the active region and are both spatially discrete and transient.
In order to constrain the possible explanations for this activity, we undertake non-potential magnetic field modelling of the active region. This modelling indicates the existence of a weakly twisted flux rope along the polarity inversion line in the region where a filament and the strong pre-flare flows are observed. 
We then discuss how these observations relate to the current models of flare triggering. We conclude that the most likely drivers of the observed activity are internal reconnection in the flux rope, early onset of the flare reconnection, or tether cutting reconnection along the filament.  

\end{abstract}

%

\end{opening}

%
\section{Introduction}
The dynamic magnetic environment of the solar atmosphere can lead to the storage of a large quantity of magnetic energy. This energy can be released via waves or magnetic reconnection, where the rapid energy release and its observational effects are manifested as a solar flare. Models of solar flare occurrence, such as the 2D CSHKP model \citep{charmichael1964,sturrock1966,Hirayama1974,kopppneuman1976} and the recent 3D model \citep{janvier2014}, describe the flaring process and associated physical effects but make little or no mention of the trigger mechanism. Recent work has identified several mechanisms and observational effects that could signify the flare trigger, but as yet a definitive trigger model has not been forthcoming.

Prior to flare onset,``free''\,magnetic energy must be built up in the active region. This energy build up can occur in a number of ways. \citet{heyvaerts1977} suggested that the emergence of magnetic flux from below the solar surface could introduce new magnetic energy into an active region. This newly emerged flux can then interact with existing magnetic field in the solar atmosphere, resulting in the triggering of solar flares (see \citeauthor{mhdofthesun} (\citeyear{mhdofthesun}) and references therein). Observational evidence for flux emergence comes through the use of magnetograms (both line-of-sight and vector). Solar flares are predominantly observed to occur at sites of strong magnetic field gradient. At these polarity inversion lines (PILs), energy can be built up in the field lines linking the two polarity regions through a shearing process. Shearing can be brought about by the rotation of sunspots\,\citep[\textit{e.g.},][]{sundararaman1998}, which can drag one polarity along the point of contact with the opposite polarity region. As rotation continues, shear increases, leading to a build-up of magnetic energy and eventually to reconnection resulting in a solar flare.  
  
  Once energy has built up in the system, several scenarios have been proposed for the initiation of large scale energy release. Tether-cutting reconnection \citep{moorelabonte1980,moore2001} is one such model of the initiation of flare and coronal mass ejection\,(CME) activity. This model proposes that slow reconnection can occur at the foot-points of a sheared loop system, resulting in the weakening of the overlying magnetic field. This weakening of the overlying field allows the filament that is supported by the sheared loop system to rise, in what is called the slow rise phase. At some point during the rise, the system becomes torus unstable and at the same time a current sheet forms beneath the rising structure. Reconnection occurs in the current sheet, accelerating the filament in the fast-rise phase and causing flaring. This relationship between tether cutting reconnection and the slow and fast rise phases of an eruption are supported by observational studies such as those by \citeauthor{chifor2006}\,(\citeyear{chifor2006,chifor2007}). In contrast to tether-cutting, \citet{antiocos1999} presented the breakout model of flare/CME triggering. This model involves reconnection occurring above a sheared arcade at a magnetic null-point created between the arcade and overlying field. This reconnection weakens the overlying field and allows the eruption of magnetic flux from the centre of the arcade. Intensity enhancements during the pre-flare period such as those observed by \citet{Warren2001} away from the site of flaring have been linked to the breakout model.
  
  The activation of the filament in an active region has also been linked to the occurrence of flaring. \citet{rubiodacosta2012} discussed how small scale reconnection within a filament leads to its destabilisation and subsequent flaring and eruption. Recent 3D  magnetohydrodynamic (MHD) simulation work by \citet{kusano2012} has identified small magnetic disturbances occurring near the PIL that may be a candidate for a flare trigger. A follow up study by \citet{bamba2013} investigated these disturbances, and the small scale internal reconnection that they cause, through observation of Ca {\sc{ii} h} emission line intensity enhancements. 
  
  Plasma instabilities are another possible way in which flaring can be triggered. \citet{torokkliem2005} used MHD simulations to propose a model of solar eruptions triggered by the helical kink instability. This instability is triggered when twist in a magnetic flux rope\,(MFR) exceeds a critical value and causes the MFR to kink and rise upwards. There have been observations that support this model, such as \citet{williams2009} who observed asymmetric Doppler shifts along a filament prior to eruption which they interpreted as a MFR subjected to the effects of the kink instability. \citet{kliemtorok2006} proposed a further plasma instability as a trigger mechanism, the torus instability. In this model, if the Lorentz force provided by an external magnetic field decreases faster than the hoop force exerted by an expanding current ring embedded in the external field, the system will be unstable. These instabilities can then rapidly cause the current ring to rise, triggering an eruption. Authors such as \citet{zuccarello2014} have provided observational evidence for the role of torus instabilities in triggering eruptions.

   The trigger mechanisms discussed thus far have clearly defined observational signatures. There are however pre-flare features that have been identified that are not yet associated with a specific flare trigger model. Non-thermal velocity\,($V_{\textrm{nt}}$) enhancements have been identified up to an hour prior to flares in X-rays by \citet{doschek1980}, and tens of minutes prior by \citet{Harra2001}. In EUV wavelengths \citet{harra2009} and \citet{wallace2010} observed V$_{\textrm{nt}}$ enhancements over an hour before flare occurrence, and \citet{harra2013} related this type of enhancement to sites of coronal dimming. They also suggested that these enhancements could be another indicator of filament activation prior to flaring.

   The X-class flare of 29 March 2014 has been a source of intense study by many authors, due to the quality and variety of data available \citep[\textit{e.g.}][]{Judge2014,Matthews2015,Li2015,Battaglia2015,Young2015,Lui2015_mgii,Aschwanden2015,KH2016,rubiodecosta2016}. There have been studies investigating aspects of the pre-flare period of this flare. \citet{abramov2015} carried out microwave and radio observations of the active region that produced the X-flare, following the evolution of the region from its appearance on disk until flaring. \citet{yang2016} used vector magnetic field extrapolations to investigate the magnetic field configuration in the active region between 28 and 29 March 2014. This work identified the presence of a magnetic flux rope in this region and suggested that the majority of the flares during their period of study were triggered by the kink instability. \citet{Kleint2015} studied the filament eruption of this flare in detail, finding that the filament exhibited accelerations of $\approx$3\,-\,5\,km\,s${^{-2}}$ during the eruption. They also identified small Doppler shifts (velocities of 2\,-- 4\,km\,s$^{-1}$) along the filament up to an hour prior to the flare observed by the \textit{Interferometric BIdimensional Spectrometer} (IBIS), which they attributed to either plasma flows or the slow rise of the filament.
    
Whilst these previous studies investigated aspects of the pre-flare behaviour of this flare, none carried out a dedicated spectroscopic investigation of pre-flare activity.  We therefore present the first spectroscopic study of pre-flare activity to simultaneously observe the chromosphere, transition region and corona using data from \textit{Hinode}/EIS and IRIS spectrometers. Section 2 provides an overview of the observations. We present the findings of these observations in Section 3, including the identification of transient, strongly blue-shifted features, characterised by transition region velocities of up to 200\,km\,s$^{-1}$ measured in emission line wings. These features are determined to be plasma flows and seen up to $\approx$40 minutes prior to flare onset.  We discuss how these features relate to extended bright features observed alongside the filament in the run up to the flare. The results of non-potential magnetic field modelling of the active region are presented, revealing the presence of a magnetic flux rope. Section 4 contains discussion of the results and how they may relate to current models of flare triggering.

\begin{figure}[]
\centerline{\includegraphics[width=1.0\textwidth,clip=]{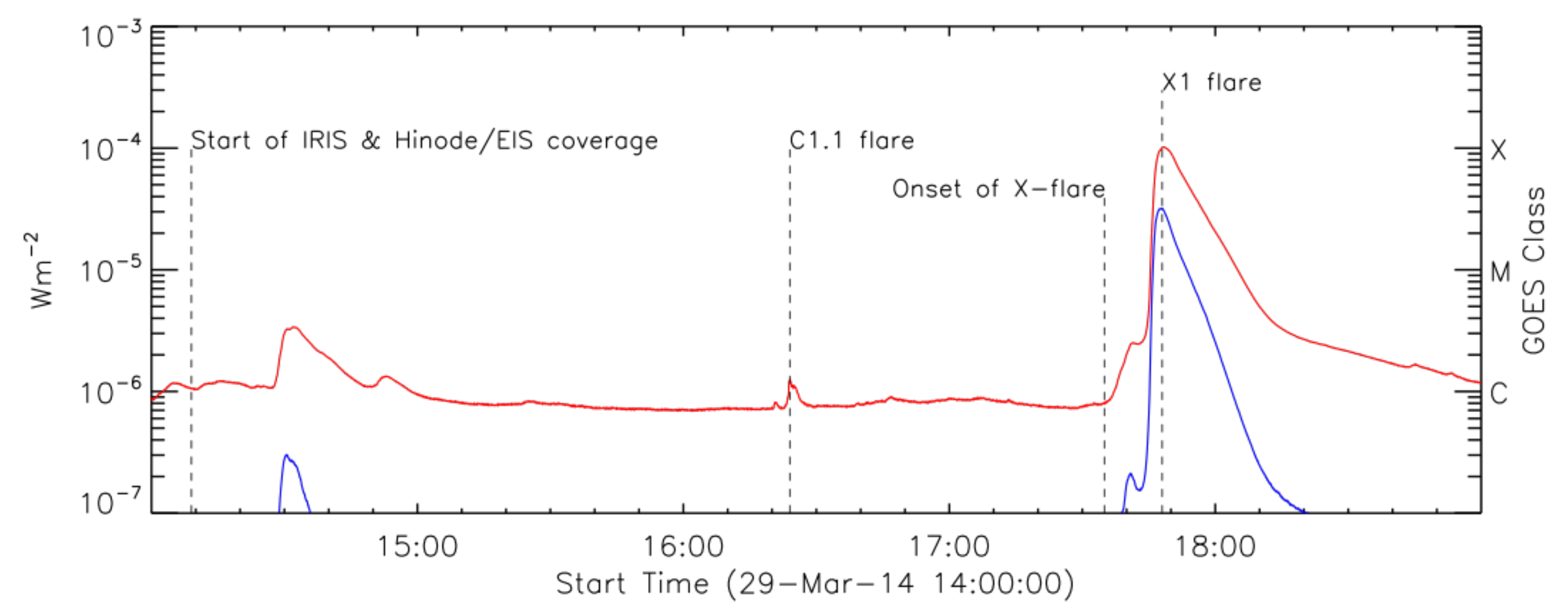}}
\caption{GOES light curve of the soft X-ray flux from 29 March 2014 14:00\,UT. Joint \textit{Hinode}/EIS and IRIS coverage starts at 14:09\,UT. The flare at 14:30\,UT occurs outside the spectrometer field of view. At 16:24\,UT a C 1.1 flare was observed by both spectrometers, followed by the X 1 flare at the peak time of 17:48\,UT.}
\label{fig: goes}
\end{figure}
\section{Observations}
\begin{figure}[]
\centering
\includegraphics[width=0.8\textwidth, clip=]{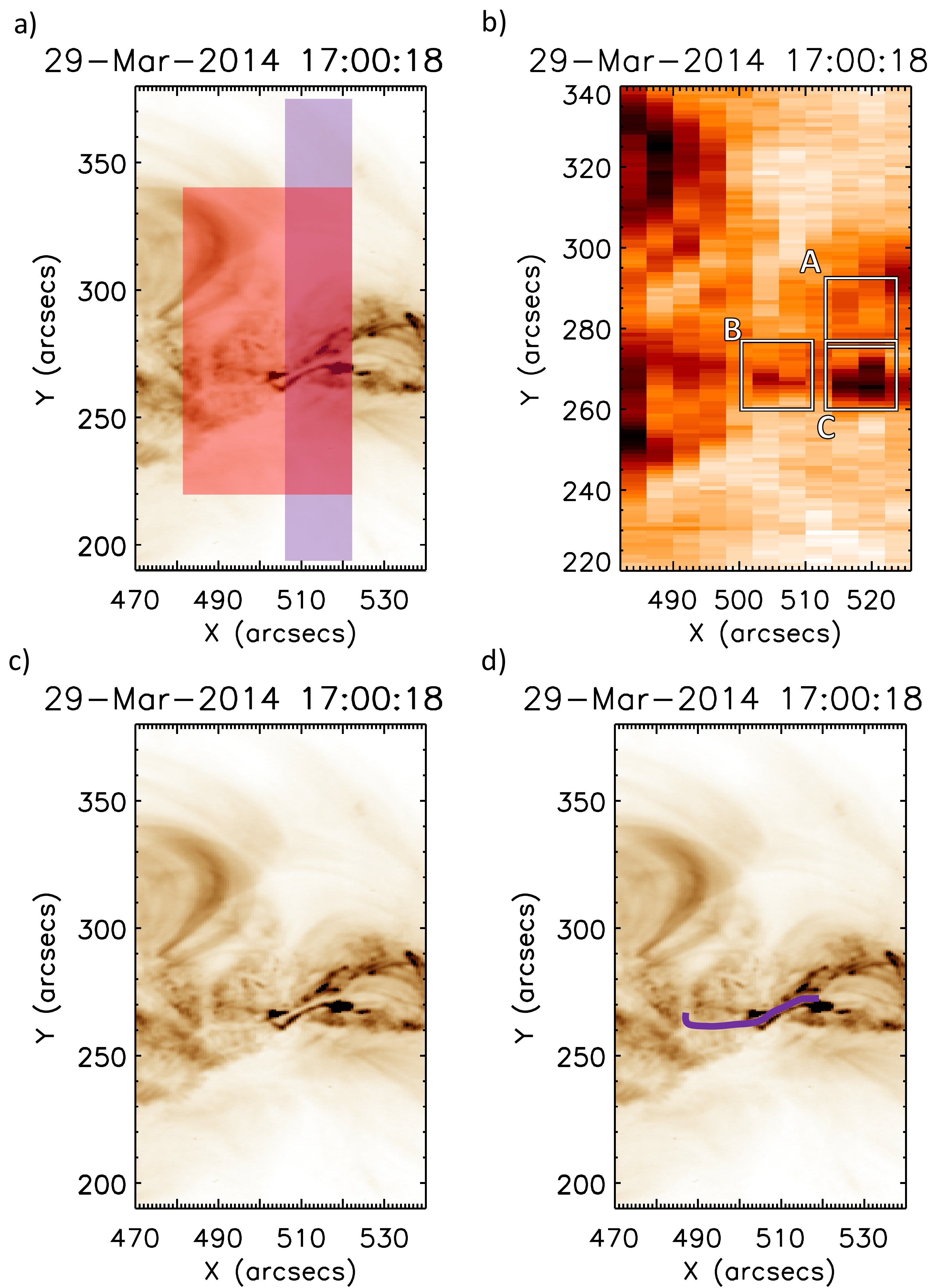}
\caption{Panel (a) shows the fields of view of the \textrm{IRIS}\,(blue) and \textit{Hinode}/textrm{EIS}\,(red) spectrometers overlayed onto the active region seen in the 193\,\AA\ AIA channel, presented with an inverted colour table. All images have been differentially rotated to the time of the closest AIA 193\,\AA\ exposure at the time of the central \textit{Hinode}/\textrm{EIS} raster step.  Panel (b) shows the three regions of study and their positions within the \textit{Hinode}/\textrm{EIS} Fe\,{\textsc{xii}} field of view. The position of the filament is clearly seen in the 193\,\AA\ and 304\,\AA\ AIA channels, with panel (c) showing the 193\,\AA\ data. The path of the filament is marked in panel (d) by the purple line for emphasis.}
\label{fig:fov}
\end{figure}  
Beginning at SOL2014-03-29T17:35, \textit{National Oceanographic and Atmospheric Administration} (NOAA) active region\,(AR) 12017 produced an X1 flare. This flare was observed by an unprecedented number of observatories, both space and ground based. Figure~\ref{fig: goes} shows the GOES soft X-ray light curve from 14:00\,UT, with the times of relevant events marked.

The \textit{Extreme Ultraviolet Imaging Spectrometer} \citep[EIS;][]{culhane07} onboard the \textit{Hinode} spacecraft \citep{kosugi07} was observing AR\,12017 continuously from 14:05 -- 17:57 UT.  During this time 104 rasters of the field of view were produced with a cadence of 134\,s. The observing program makes use of the 2$''$ slit rastering across a field of view of 42$''$\,$\times$\,120$''$ in 4$''$ steps (shown in red in Figure~\ref{fig:fov}, panel (a)). Eight spectral windows are contained within the observing program, of which Fe\,{\sc{xii}} 192.39\AA\ and He\,{\sc{ii}} 256.28\AA\ are selected for analysis. Fe\,{\sc{xii}} is selected as it is the strongest coronal line observed by EIS providing the best opportunity of identifying low intensity pre-flare activity. He\,{\sc{ii}} was chosen to study the lower atmosphere due to its pseudo-chromospheric nature.

 The \textit{Interface Region Imaging Spectrometer} \citep[IRIS;][]{depontieu14} was observing AR\,12017 from 14:09 -- 17:54 UT with a field of view of 14$''$\,$\times$\,174$''$, produced using an eight step raster of 2$''$ steps (highlighted in blue in Figure~\ref{fig:fov}, panel (a)). Although nominal exposures were 8 seconds, automatic exposure control was in effect during rasters covering the peak of the X-flare, reducing the exposure time to $\approx$2 seconds. These exposure times result in a raster cadence of 72\,s. \textrm{IRIS} also utilises a slit jaw imager\,(SJI) to provide context to spectroscopic observations.

Data from NASAs \textit{Solar Dynamics Observatory} \citep[SDO;][]{pensell12} \textit{Atmospheric Imaging Assembly} (AIA) and \textit{Helioseismic and Magnetic Imager} (HMI) instruments were downloaded with calibrations and corrections already applied. The routine \texttt{aia\_prep} was run on the data before use to account for small scaling differences between the full disk images produced by both instruments.

In order to directly compare spatial locations between the four data sets used in this analysis, \textit{Hinode}/EIS and IRIS images are aligned with AIA images. This alignment is done manually using feature recognition of the corresponding AIA wavelength to the chosen \textit{Hinode}/EIS or IRIS data. To do this, \textit{Hinode}/EIS Fe {\sc{xii}} 192.6\,\AA\ data are aligned to AIA\,193\,\AA\ data for each time step, while for the IRIS alignment, Si {\sc{iv}} 1400\,\AA\ observations are aligned to the AIA 1600\,\AA\ channel. The alignment by feature recognition requires only very small changes to position, of no more than $\pm$2 arcseconds for each individual raster.

\section{Results}
\subsection{Active Region Evolution}
\begin{figure}[]
\centerline{\includegraphics[width=1.0\textwidth,clip=]{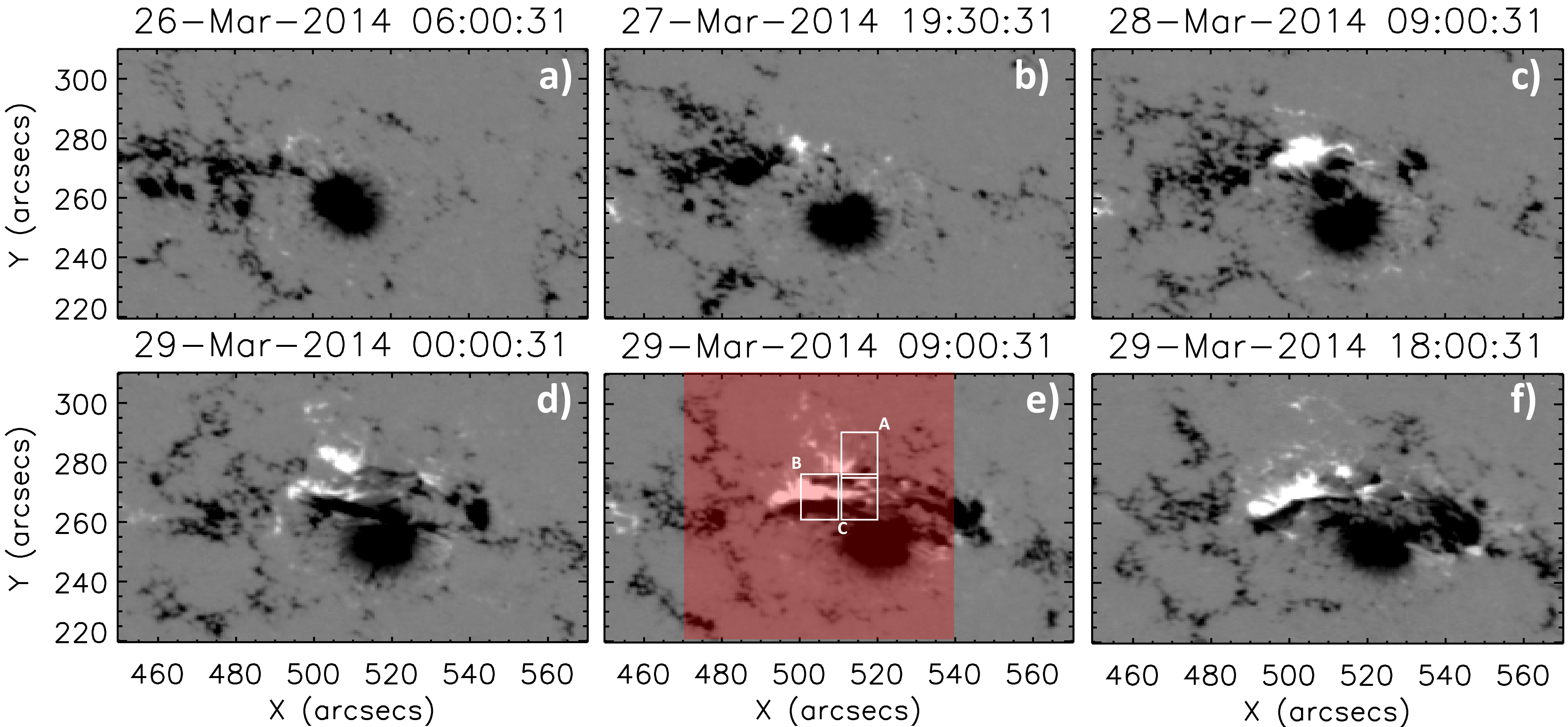}}
\caption{Evolution of the central portion of AR 12017 between 26 March 2014 and 29 March 2014. Flux emergence is observed in panel (b), and subsequent panels show the formation of a highly sheared polarity inversion line. Plots are scaled between $\pm$550 Gauss. Panel (e) also indicates the AIA field of view used in Figure \ref{fig:fov}, marked by the red shaded area, as well as the three regions of study.}
\label{fig:hmievo}
\end{figure}

AR 12017 was first observed crossing the eastern limb of the Sun on 22-March-2014. The evolution of this active region from 22 March 2014 to 29 March 2014 was discussed in \cite{abramov2015}. They showed that there is a large negative polarity sunspot leading the AR with smaller less coherent positive polarity trailing until 27 March 2014 when a new positive polarity region emerges next to the leading negative polarity. This is illustrated in Figure~\ref{fig:hmievo}, which shows the evolution of the central portion of the active region. Panel (a) displays the initial morphology, with the clear large negative sunspot. In panel (b), we observe the emergence of the positive polarity region. Panels (c)\,--\,(f) show the evolution of this emergence and its subsequent interaction with the existing negative polarity. It is interesting to note that as this interaction progresses, the apparent motion of the positive polarity with respect to the polarity inversion line\,(PIL) indicates an increase in shear of the magnetic field. Panel (f) shows the AR just after the peak of the X-class flare. We see that the field has become sheared over the PIL, suggesting that there has been a large build up of magnetic energy in this region.

Within the AR on 29 March 2014, there is a clearly visible filament.  This feature can be seen clearly in the AIA 193\,\AA\ observations in Figure\,\ref{fig:fov}, panels (c) and (d). The path of the filament can also be clearly seen in the H$\alpha$ observations presented in Figure 5 of \cite{Kleint2015}.

\subsection{Pre-Flare Observations in the Corona}
Previous studies of pre-flare behaviour have found enhanced $V_{\textrm{nt}}$ features observed in the corona over periods of tens of minutes to an hour prior to flare onset, suggesting increased V$_{\textrm{nt}}$ may be an indicator of imminent flare onset \citep{harra2009}. In order to investigate whether this phenomenon was observed prior to the 29 March 2014 X-flare, Fe\,\textsc{xii} 192.3\,\AA\ observations from \textit{Hinode}/EIS are analysed. These data are fitted with single Gaussian profiles. Due to the lack of absolute wavelength calibration in \textit{Hinode}/EIS data, care is taken to determine a rest wavelength. This is done by selecting a small area of the field of view from a pre-flare raster, in which little activity is observed and fitting to determine the rest wavelength. The raster chosen for this purpose is the first in the data set, recorded at 15:01:09\,UT. Values for Doppler velocity and $V_{\textrm{nt}}$ are determined from the results of the spectral fitting. Doppler velocity is determined by measuring the deviation of line centre from the rest velocity, while $V_{\textrm{nt}}$ is defined as the width of a line observed above that of the theoretical thermal line width for the given ion under observation.

\begin{figure}[]
\centerline{\includegraphics[width=1.0\textwidth,clip=]{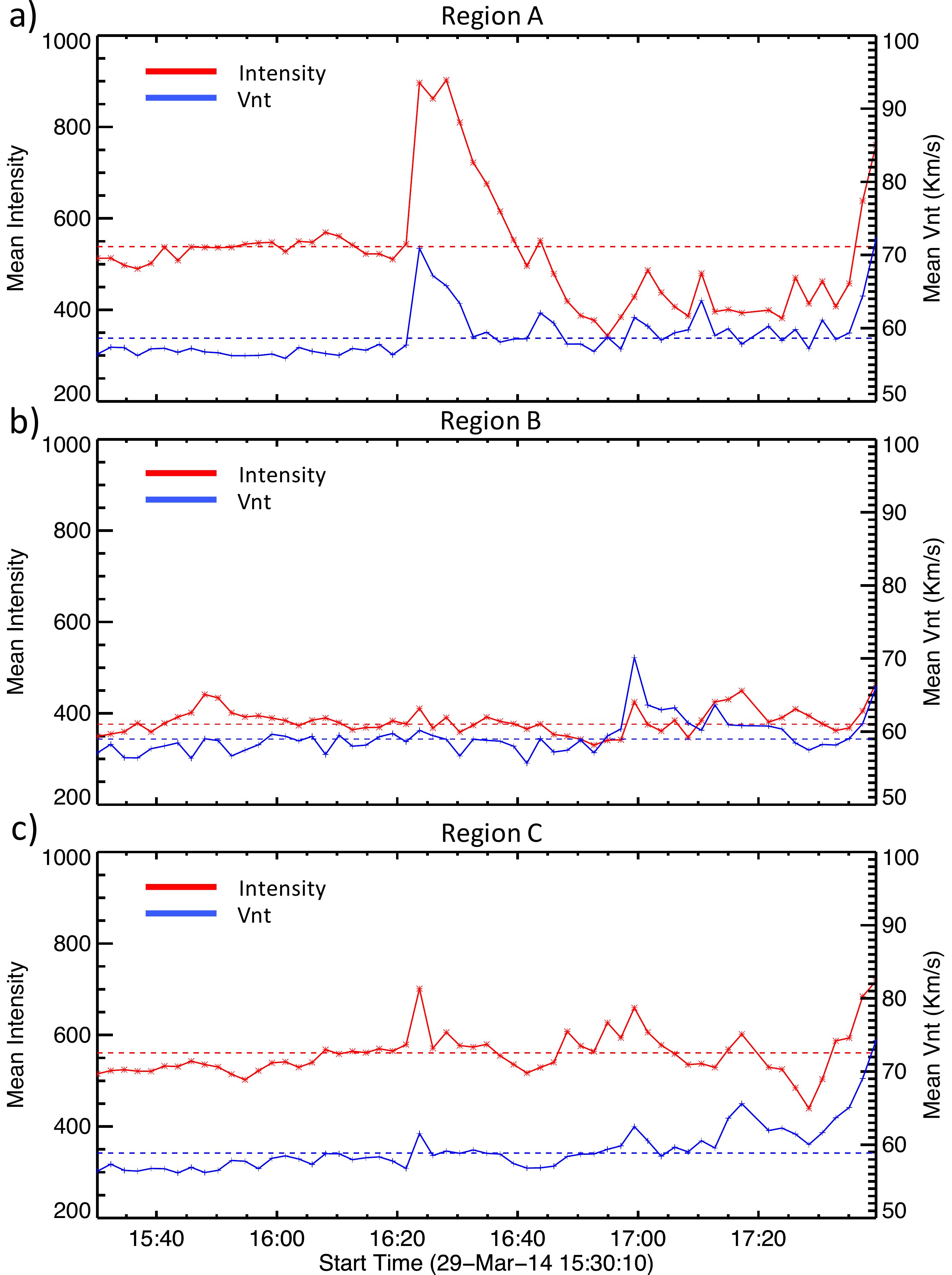}}
\caption{Evolution of mean intensity\,(red) and mean $V_{\textrm{nt}}$\,(blue) calculated for Fe\,{\sc{xii}}, in time for Regions A, B and C, as shown in Figure~\ref{fig:fov}. In Region A, shown in panel (a), the two profiles broadly match, in particular showing a clear peak during the 16:24\,UT C1.1 flare. The location of $V_{\textrm{nt}}$ enhancement during the C-flare is shown in Figure~\ref{fig:aiavnt} panel (a).
 In Region B, panel (b), we can see that at 17:00\,UT, non-thermal velocity peaks at a value of $\approx$70\,km\,s$^{-1}$. This non-thermal velocity peak occurs in the absence of a corresponding intensity increase. The location of this region of $V_{\textrm{nt}}$ enhancement is shown in Figure~\ref{fig:aiavnt} panel (b).
 Region C, panel (c): both intensity and non-thermal velocity profiles match well, with no irregularities as in the case of Region B. A smaller response to the C1.1 flare is observed in this region. Prior to 17:00\,UT increases in intensity and non-thermal velocity are observed. Region C also exhibits the earliest response to the X flare of the three regions studied. Figure~\ref{fig:aiavnt} panel (c) shows the location of non-thermal velocity enhancements}
 \label{fig:lightcurves}
\end{figure}

\begin{figure}[]
\centerline{\includegraphics[width=0.8\textwidth,clip=]{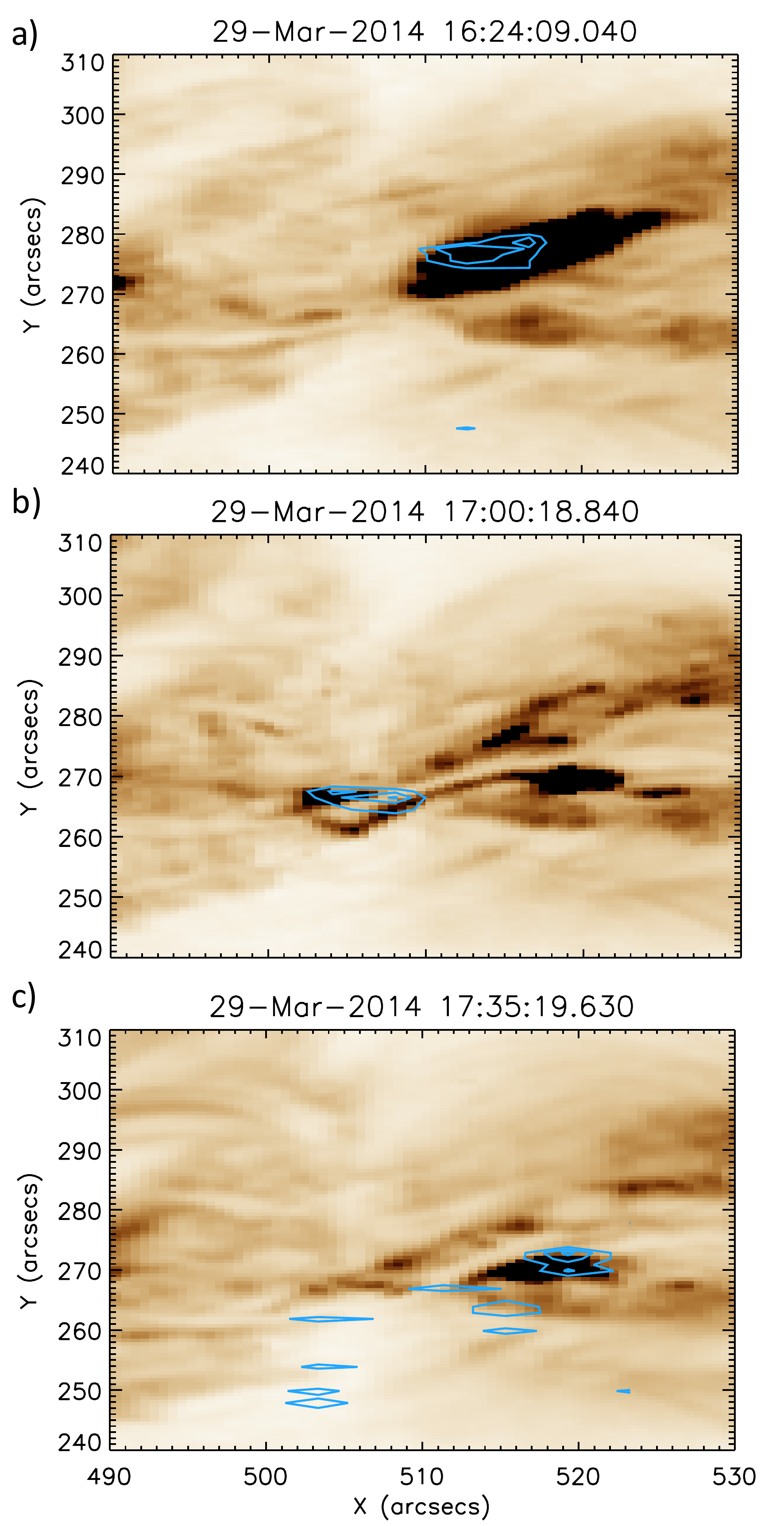}}
\caption{This figure shows non-thermal velocity contours overlayed onto corresponding AIA 193\,\AA\ images, to exhibit the location of the non-thermal velocity enhancements. $V_{\textrm{nt}}$ contours are plotted between 70 and 200 km\,s$^{-1}$, and the AIA color scale is inverted to improve clarity. Panel (a) shows the location of these enhancements during the C-flare at 16:24\,UT. Panel (b) shows that the $V_{\textrm{nt}}$ enhancements seen at 17:00\,UT are located in a region in the center of the filament. Enhancements seen at the start on the X-flare, at 17:35\,UT, are shown in panel (c). }
 \label{fig:aiavnt}
\end{figure}

Figure~\ref{fig:lightcurves} shows time profiles of mean intensity (red) and mean $V_{\textrm{nt}}$ (blue) for three chosen subregions: A, B and C. This method of investigating the behaviour using $V_{\textrm{nt}}$ averaged over a small region is similar to that employed in earlier coronal studies of pre-flare activity \citep[\textit{e.g.}][]{harra2009, harra2013}. Region A is chosen as it is the site of the confined C-class flare that occurred at 16:24\,UT. From Figure~\ref{fig:lightcurves} panel (a), we see that the intensity and $V_{\textrm{nt}}$ time profiles match closely, both exhibiting a strong peak during the C-class flare. This region of V$_{\textrm{nt}}$ enhancement is clearly seen in panel (a) of  Figure\,\ref{fig:aiavnt}, where Fe\,{\sc{xii}} V$_{\textrm{nt}}$ contours are overlayed onto AIA 193\,\AA\ images at the peak of the C-flare. The time profile of Region B (Figure~\ref{fig:lightcurves}, panel (b)) shows little activity early in the observations as this region is uninvolved in the C-flare at 16:24\,UT. However, at $\approx$17:00\,UT there is a clear spike in the mean $V_{\textrm{nt}}$ for Region B, accompanied by a small increase above the mean intensity. The peak mean $V_{\textrm{nt}}$ observed in Region B is over 70\,km\,s$^{-1}$. This is a significant value for an active region at a time when no flaring activity is occurring. \citet{testa2016} have found $V_{\textrm{nt}}$ values in non-flaring active regions to be 24\,km\,s$^{-1}$, as determined from the Fe\,{\sc{xii}} 1349.4\,\AA\ emission line. It can be clearly seen in panel (b) of Figure\,\ref{fig:aiavnt} that this small section of $V_{\textrm{nt}}$ enhancement is located directly over the area of the active region where the filament is situated. From the Doppler velocity data, this feature ia identified as having a line-centre blue-shift of tens of km\,s$^{-1}$.
 The time profile of Region C (Figure~\ref{fig:lightcurves}, panel (c)) is again different to that seen in the other regions. At 16:24\,UT there is a small response to the C 1.1 flare seen in both intensity and $V_{\textrm{nt}}$. This lesser response to the C-flare in this region results from the effects of the flare only being present in a small subregion which is common to both Regions A and C. Region C also shows intriguing activity from $\approx$16:45\,UT to 17:00\,UT, both in intensity and $V_{\textrm{nt}}$. The intensity profile during this period exhibits three periodic peaks, each with a higher peak intensity than the previous. This periodic increase in intensity is accompanied by a general increase in $V_{\textrm{nt}}$. At $\approx$17:17\,UT there is a peak in intensity and $V_{\textrm{nt}}$. The onset of the X-class flare is seen in Region C from $\approx$17:28\,UT with clear and sustained increases in intensity and $V_{\textrm{nt}}$. These effects of the X-flare are seen in Region C up to five minutes prior to similar effects in Regions A and B. V$_{\textrm{nt}}$ enhancements at the start of the X-flare (17:35\,UT) can be seen in panel (c) of Figure\,\ref{fig:aiavnt}.
Having found sources of enhanced coronal $V_{\textrm{nt}}$, we explore the behaviour of the lower atmosphere at the pixel scale.

\subsection{Response of the lower atmosphere}
The response of the lower atmosphere in these regions is also investigated to fully understand pre-flare activity throughout the solar atmosphere. The transition region response is investigated via the Si {\sc{iv}} line at 1402.77\,\AA\ observed by IRIS. The pseudo-chromospheric He {\sc{ii}} 256.2\,\AA\, observed by EIS and the optically thick chromospheric Mg {\sc{ii}} h and k lines (at 2803.52\,\AA\ and 2796.34\,\AA\ respectively) observed by IRIS are chosen to investigate the atmosphere to chromospheric depths. To investigate the dynamics in these regions, the evolution of the profiles of each spectral line is studied between 16:16\,UT and 18:00\,UT. The data for each line profile sequence are plotted as a function of time and Doppler velocity, centred on the rest velocity of each individual line. These rest velocities are calculated by fitting a small, inactive area of a pre-flare raster. The following is an account of the pre-flare activity observed for each sub-region in turn.

\subsubsection{Region A: Site of C-class Flare}
\begin{figure}[]
\centerline{\includegraphics[width=1.2\textwidth,clip=]{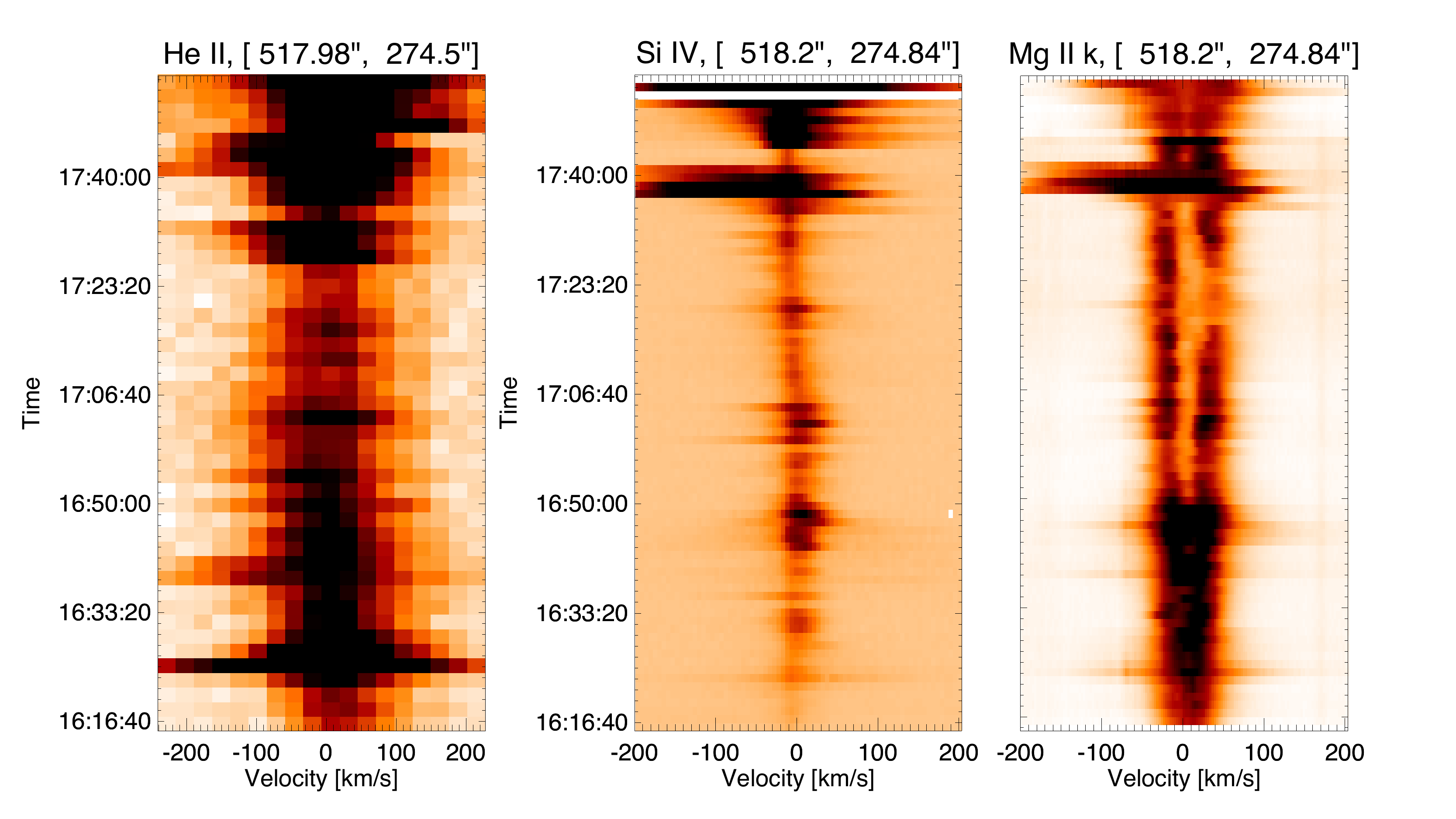}}
\caption{Spectral time profiles recorded in Region A at 16:16\,UT. Panel (a) shows He\,{\sc{ii}} spectra; panel (b) shows Si\,{\sc{iv}} spectra; and panel (c) shows the evolution of the Mg\,{\sc{ii}}\,k line during this time period. Broadening of the lines is observed at 16:24\,UT in response to the C-flare. The response to the X-class flare can be seen in all lines from $\approx$17:35\,UT.}
\label{fig:rega}
\end{figure}

 In Region A (Figure~\ref{fig:lightcurves}, panel (a)), the intensity and $V_{\textrm{nt}}$ profiles track each other closely, both exhibiting a strong peak during the C1.1 flare at 16:24\,UT. 
Figure~\ref{fig:rega} illustrates the He {\sc{ii}}, Si {\sc{iv}} and Mg {\sc{ii}} k line profile evolution for Region A. We clearly see in the He {\sc{ii}} spectra the increase in intensity and line width due to the C1.1 flare at 16:24\,UT, as well as lesser dynamics in both lines up until the onset of the X-flare at 17:35. Both He\,{\sc{ii}} and Si\,{\sc{iv}} spectra show a strongly blue-shifted feature at $\approx$17:40\,UT, which we interpret as the eruption of the filament during the X-flare. The Mg {\sc{ii}} k emission is observed to have strong intensity and a weak central reversal in its profile from the onset of the C-flare until $\approx$16:50\,UT. At this point in time the central reversal deepens and the line intensity is weaker on the whole. As in the case of He {\sc{ii}} and Si {\sc{iv}}, strong blue asymmetries are seen in the Mg {\sc{ii}} data at $\approx$17:40\,UT.
Thus within Region A, we observe a similar response to the C-flare throughout the atmosphere, noting that the activity is in line with that expected for an area of study such as this.

\subsubsection{Region B: Site of Pre-Flare $V_{\textrm{nt}}$ Enhancement}
Region B is chosen for further study due to an intriguing $V_{\textrm{nt}}$ feature observed at 17:00\,UT.
\begin{figure}[]
\centerline{\includegraphics[width=1.2\textwidth,clip=]{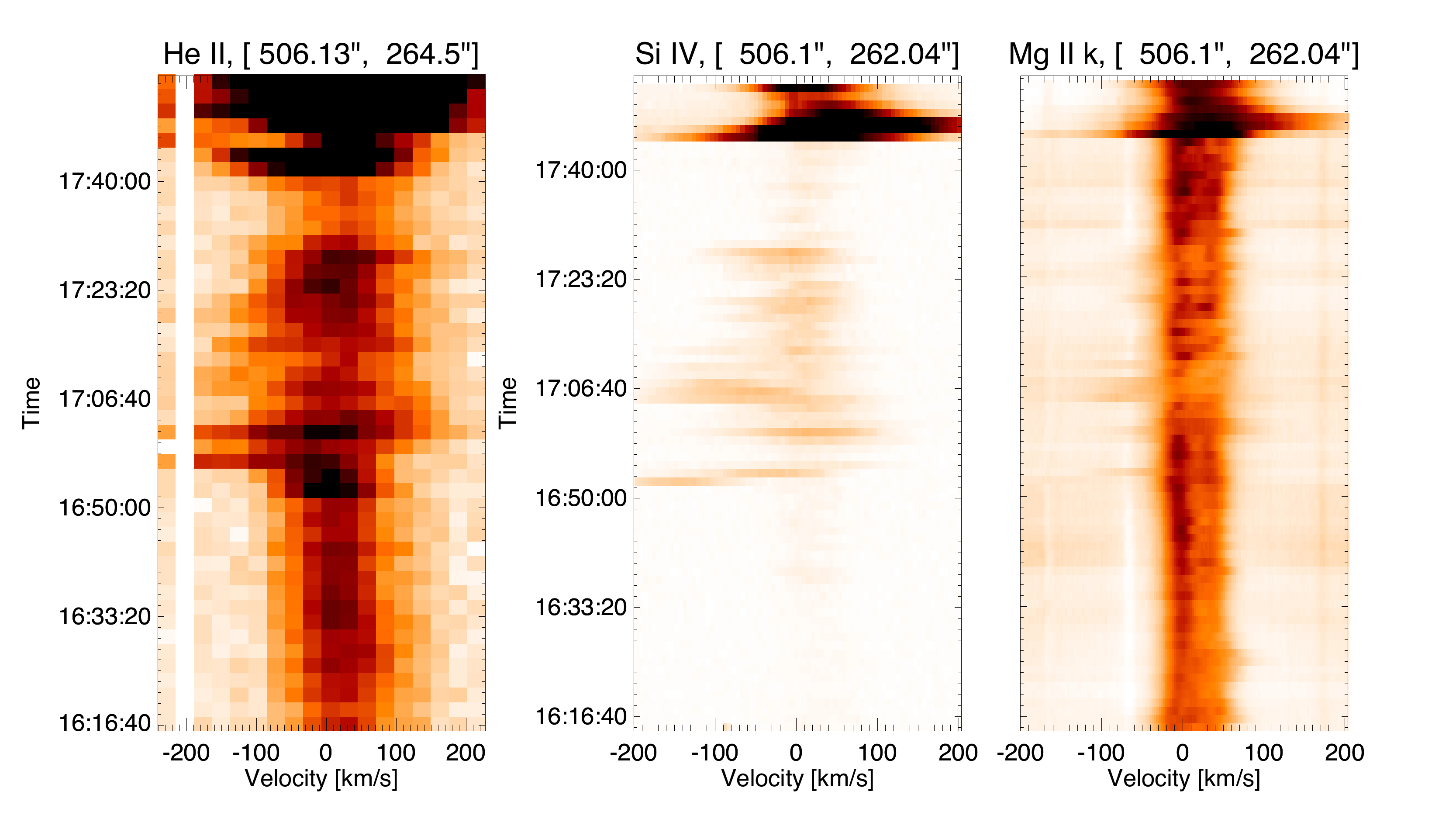}}
\caption{Spectral time profiles recorded in Region B at 16:16\,UT. Panel (a) shows He\,{\sc{ii}} spectra; panel (b) shows Si\,{\sc{iv}} spectra; and panel (c) shows the evolution of the Mg\,{\sc{ii}}\,k line during this time period. Little activity is observed until the onset of strong blue shifts are seen from $\approx$16:52\,UT. The onset of the flare can be seen to occur in this region from 16:40\,UT.  }
\label{fig:regb}
\end{figure}
Figure~\ref{fig:regb} shows the evolution of the He {\sc{ii}}, Si {\sc{iv}} and Mg {\sc{ii}} k line profiles in Region B. As expected there is no observed activity during the C1.1 flare as this is confined solely to Region A. 
From the start of observation until $\approx$16:50\,UT, the spectra show ordinary active region dynamics with little to no line broadening or Doppler shifts.
  From $\approx$16:50\,UT however, strong blue shifts of up to 200\,km\,s$^{-1}$ are observed to initiate very rapidly in both He\,{\sc{ii}} and Si\,{\sc{iv}} lines. Additionally, these strong blue shifts are observed to be very dynamic, reaching a maximum around 16:54\,UT, falling at 17:00\,UT, and then reaching 200\,km\,s$^{-1}$ from 17:04\,UT. This behaviour is very interesting as peak coronal blue shifts, determined from Fe\,{\sc{xii}} data, are observed at 17:00\,UT, suggesting there is a temporal offset between the layers of the atmosphere involved in the activity. After these peak blue shifts, the line profiles continue to be dynamic until the onset of the X-flare, albeit exhibiting blue shifts of lesser velocities of up to  $\approx$100\,km\,s$^{-1}$. 
  Within this region, we identify increased $V_{\textrm{nt}}$ from coronal Fe\,{\sc{xii}} data, as well as strong blue-shifted flows in the lower atmosphere which initiate very rapidly. 

\subsubsection{Region C: Site of Pre-Flare Brightening and Earliest Response to X-flare}
Region C is chosen as it is the site of a pre-flare brightening observed in AIA 193\,\AA\ data.
\begin{figure}[]
\centerline{\includegraphics[width=1.2\textwidth,clip=]{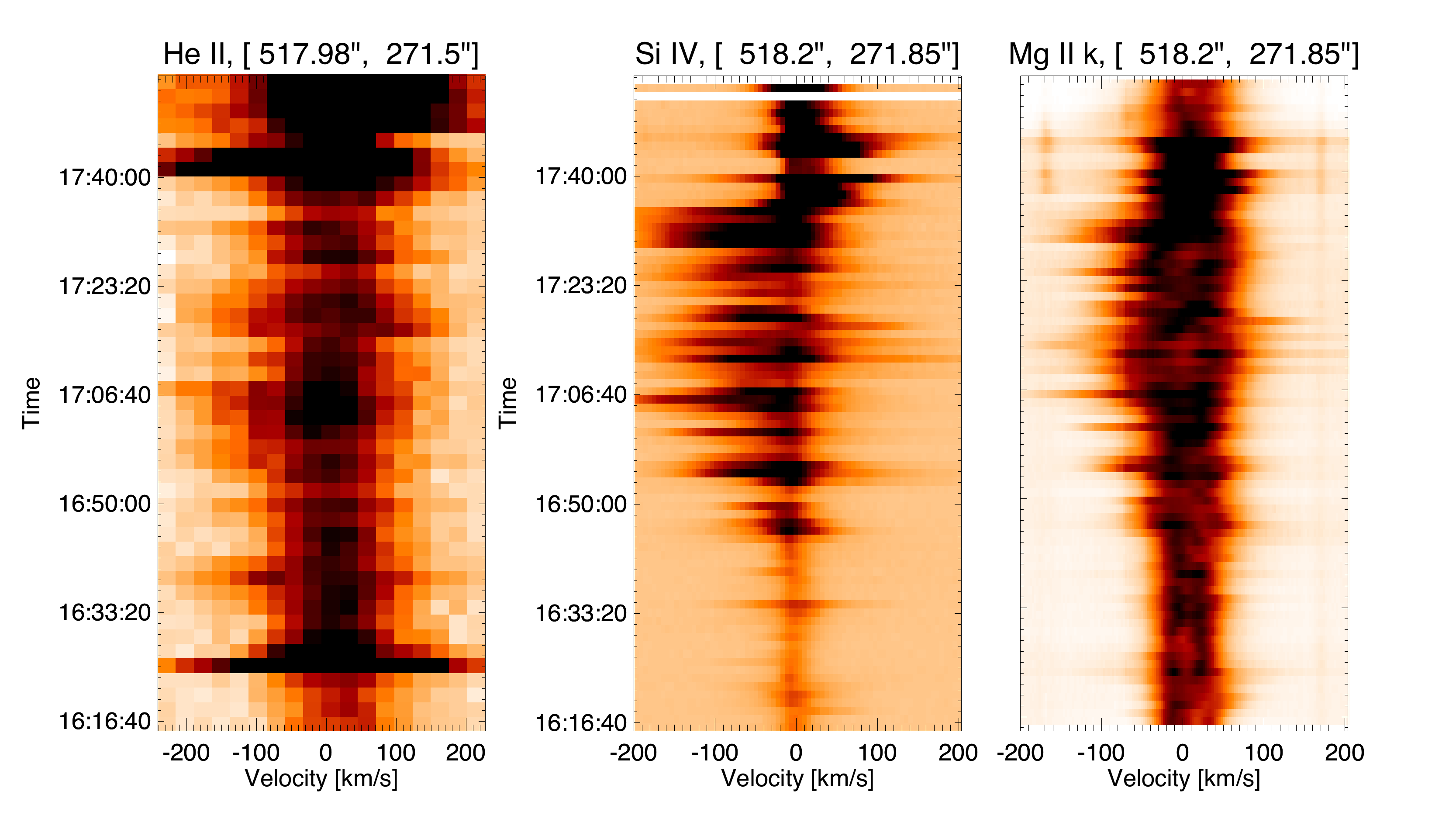}}
\caption{Spectral time profiles recorded in Region C at 16:16\,UT. Panel (a) shows He\,{\sc{ii}} spectra; panel (b) shows Si\,{\sc{iv}} spectra; and panel (c) shows the evolution of the Mg\,{\sc{ii}}\,k line during this time period. A small response to the 16:24\,UT C-flare is observed in this region. From 16:45\,UT until the onset of the X-flare, blue-shifted line broadening is observed.}
\label{fig:regc}
\end{figure}
The single pixel time profiles obtained in Region C, Figure~\ref{fig:regc}, differ again from those observed in Regions A and B. Little activity or line broadening is observed in Si {\sc{iv}} and Mg {\sc{ii}} k spectra prior to $\approx$16:45\,UT. The He {\sc{ii}} time profile shows a response to the C-flare at 16:24\,UT and in general shows more activity than the other lines studied during this time period. From 16:45\,UT until the onset of the X-class flare at 17:35\,UT, this activity is characterised by increased line intensity, especially in Si {\sc{iv}} observations, as well as strong, intermittent blue asymmetries in all observed lines. These blue asymmetries have speeds between 100\,km\,s$^{-1}$ and 200\,km\,s$^{-1}$.
The onset of pre-flare activity in this region is not as dramatic as that in Region B, but dynamic blue shifts of up to 200\,km\,s$^{-1}$ are observed.

\subsection{Location of Observed Plasma Flows}
\begin{figure}[]
\centering
\includegraphics[width=1.0\textwidth]{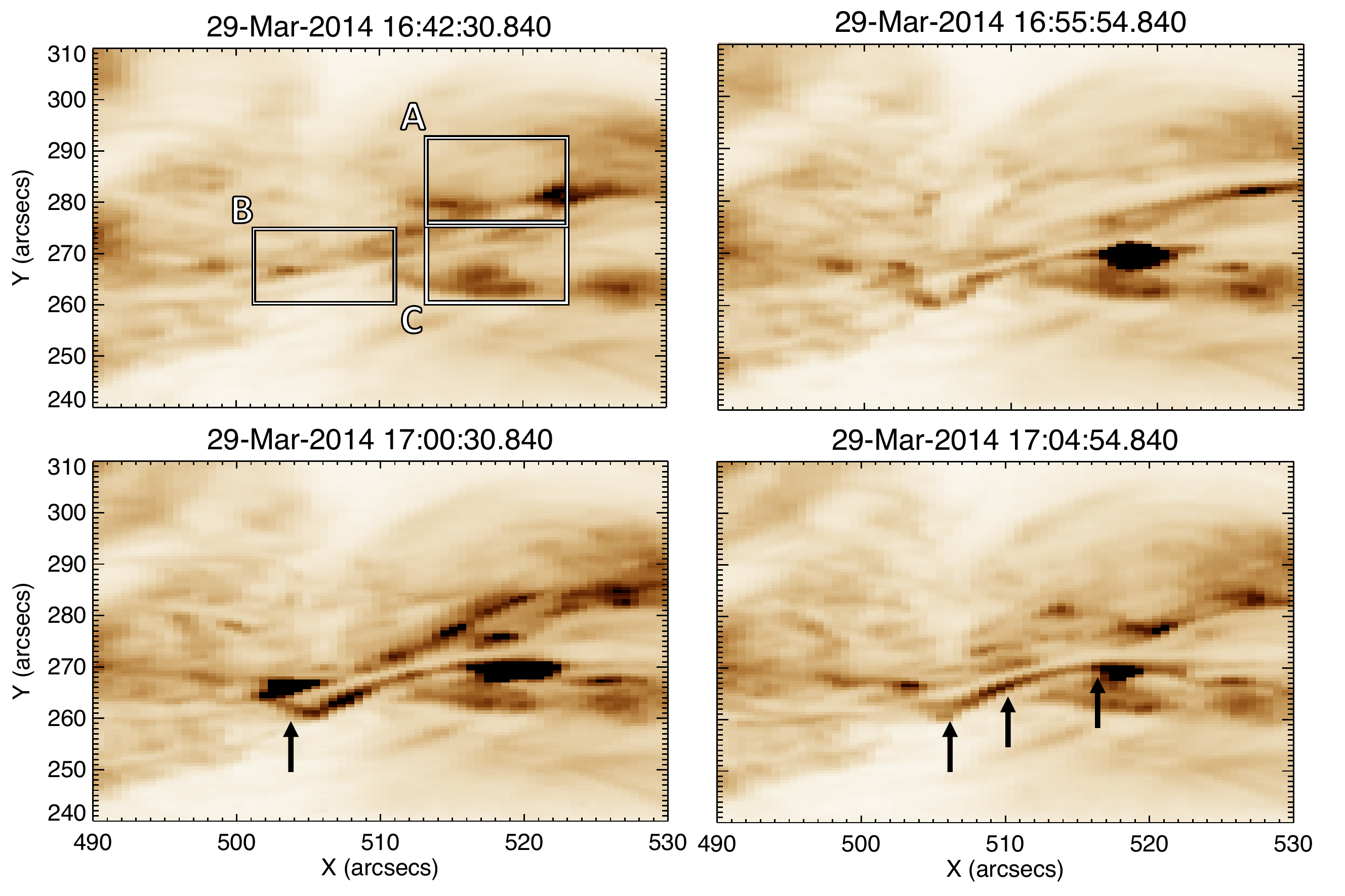}
\caption{Stills detailing coronal activity observed by SDO/AIA in 193\,\AA, shown using an inverted colour table, between 16:42\,UT and 17:04\,UT. Panel (a) shows the filament prior to any observed activity. Overlayed are the three sub-regions of study. Panel (b) shows the appearance  of a bright feature to the west of the filament. Panel (c) shows a loop feature crossing the filament at the site of the peak coronal blue shift as observed by \textit{Hinode}/EIS. This activity is marked by the black arrow in the diagram. Panel (d) shows an extended bright feature lying along the southern edge of filament. The path of this brightening is marked by the overlayed black arrows.}
\label{fig:aiacomp}
\end{figure}

\begin{figure}[]
\centering
\includegraphics[width=1.0\textwidth]{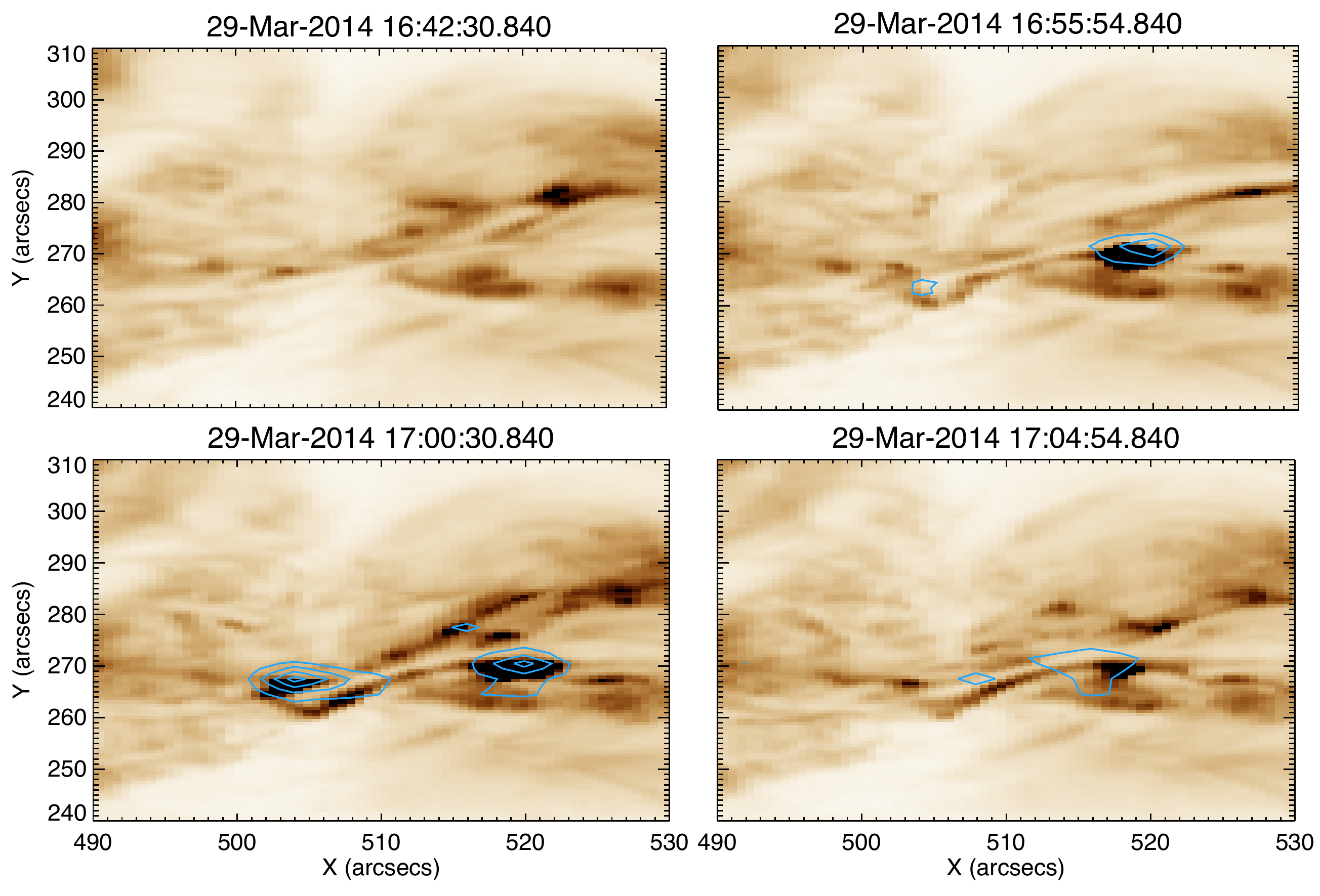}
\caption{Here we see the same AIA 193\,\AA\ (colour table inverted) field of view as Figure~\ref{fig:aiacomp}, over-plotted with \textit{Hinode}/EIS Fe\,{\sc{xii}} data at $-$100\,km\,s$^{-1}$. No strong blue shifts can be seen in the EIS data in panel (a) at 16:42\,UT. By 16:55\,UT (panel (b)) we can see that there is a strong area of blue shift located over a bright region in the AIA data. There is also a less intense region of blue shift located in the centre of the filament. At 17:00\,UT (panel (c)) both these blue-shifted regions have increased in intensity. Panel (b) shows the situation at 17:04\,UT where the strong blue shifts have dissipated greatly with the feature in the centre of the filament being no longer visible.}
\label{fig:aiaeis}
\end{figure}

From examination of the line profiles we determine that from $\approx$16:50\,UT the plasma is highly dynamic. As these line profiles represent the evolution of the spectra at one pixel in time, we must also consider the morphology of these flows and their evolution over the whole field of view. Figure~\ref{fig:aiacomp} shows four AIA 193\,\AA\ images charting the activity along the filament. In panel (a) we see the filament at 16:42\,UT and note that there is little activity seen in the areas corresponding to Regions A, B and C\,(positions are shown in the marked boxes). Panel (b) shows a bright feature appearing within Region C, lying directly to the south of the filament, at around 16:54\,UT. By 17:00\,UT (panel (c); the time of peak coronal blue shift as measured by \textit{Hinode}/EIS), we can see that a bright loop now lies across the filament, within Region B. This loop is situated directly above the site of the blue shifts observed by \textit{Hinode}/EIS and IRIS. Additionally, in both panels (c) and (d) an elongated bright feature extends from this loop west to the site of the initial brightening observed in panel (b). The path of this feature is marked by the black arrows in panel (d). 
In Figure~\ref{fig:aiaeis} we show for similar time selections the Fe {\sc{xii}} data at $-$100\,km\,s$^{-1}$ overlaid onto the AIA data. We see from panel (a) that there is no activity at this velocity in the Fe\,{\sc{xii}} data at 16:42\,UT. An area of strong blue shift is seen to appear in the region of increased intensity at 16:54\,UT, along with a weaker blue-shifted feature in the centre of the filament, shown in panel (b). Panel (c) shows that at 17:00\,UT this blue-shifted region is still present, along with a second strongly blue-shifted region in the centre of the filament. Panel (d) of Figure~\ref{fig:aiaeis} shows that at 17:04:54\,UT the blue shifts in the corona seem to extend from the two regions identified earlier, along the extended bright feature. 
\begin{figure}[]
\centering
\includegraphics[width=1.0\textwidth]{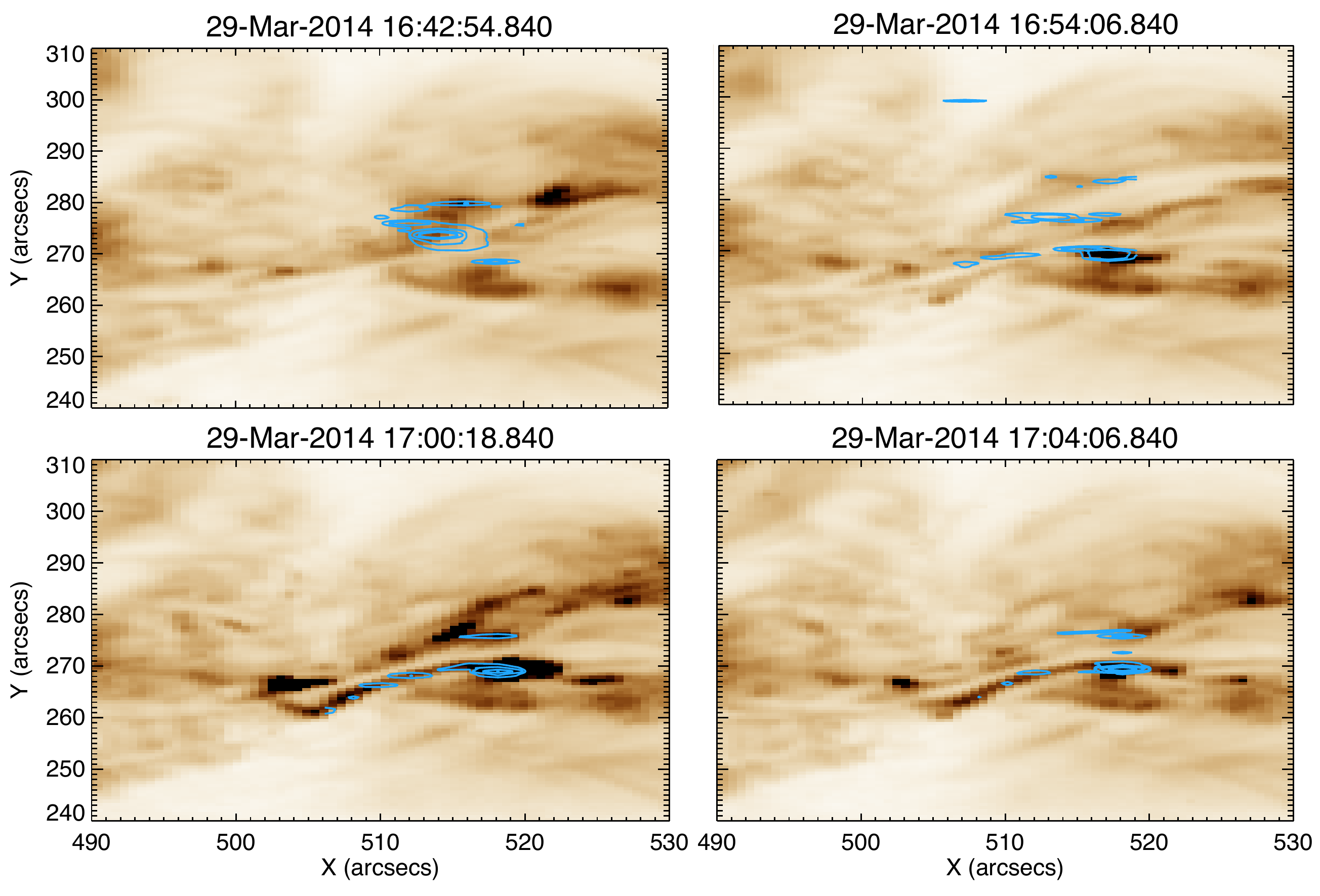}
\caption{For the same region detailed in Figures 8 and 9, this Figure displays Si\,{\sc{iv}} emission at $-$100\,km\,s$^{-1}$ observed by IRIS, overlayed onto AIA 193\,\AA\ data (colour table inverted). In panel (a), at 16:42\,UT, we see a blue-shifted region centred on the region of the earlier C-class flare. In panel (b), 16:54\,UT, blue shifts in the region of the C-flare have waned in intensity. In the region of brightening in the AIA data there is a region of strong blue shift, as well as discrete areas of blue shift extending along the filament. At 17:00\,UT, panel (c), these discrete blue shift features are located along the extended bright feature visible in AIA data. By 17:04\,UT, panel (d), blue shifts have decreased in intensity, but are still present along the extended bright feature.}
\label{fig:aiairis}
\end{figure}
Figure~\ref{fig:aiairis} details the morphology of the $-$100\,km\,s$^{-1}$ blue shifts observed in the Si {\sc{iv}} data, overlaid onto the AIA 193\,\AA\ images. Panel (a) reveals that at 16:42\,UT, blue shifts in the region of the earlier C-class flare. At 16:54\,UT (panel (b)), blue shifts are still present at the site of the C-class flare but to a lesser extent. In the region of the bright feature in the AIA 193\,\AA\ image, strong blue shifts are also observed. These blue shifts are in the same positions as those identified during the same time period from Fe {\sc{xii}} data. Additional blue shifts are also observed along the filament at this point in time.
In panel (c), we find that the blue shifts associated with the bright feature are still present. Blue shifts are also observed to lie along the bright ribbon-like feature to the south of the filament. It is noted that these blue shifts are spatially discrete. The situation seen in panel (d) at 17:04\,UT is similar to that observed at 17:00\,UT. However, the positions of the spatially discrete blue shifts along the extended bright feature have changed, highlighting the transient nature of these features. This behaviour is seen more clearly in the movie of this time period (Movie 1) in the online version. 

The fast plasma flows and brightenings, observed throughout the solar atmosphere and discussed in the preceding sections are clear examples of pre-flare activity. However several differing models could be used to explain such activity. In order to attempt to confine the possible drivers, non-potential magnetic field modelling is carried out on the active region. This work and its results are described in Section 3.5. 

\subsection{Non-potential Magnetic Field Modelling}
To simulate the non-potential evolution of the active region magnetic field and the morphology of the magnetic field at the filament location, a continuous time-series of quasi-static, nonlinear force-free fields are produced. These non-potential magnetic fields are produced using the technique developed and applied in \cite{Mackay2011} and \cite{Gibb2014}. In this technique, the  boundary driving at the level of the photosphere is obtained directly from a time series of HMI line-of-sight magnetograms which are directly applied as lower boundary conditions. The coronal magnetic field then responds to these motions by evolving through a continuous series of quasi-static nonlinear force-free fields using the magneto-frictional relaxation method.  Full details of the equations solved and the technical details of applying this technique to magnetogram data are described in the cited papers.

\begin{figure}[]
 \centering
 \includegraphics[scale=0.6]{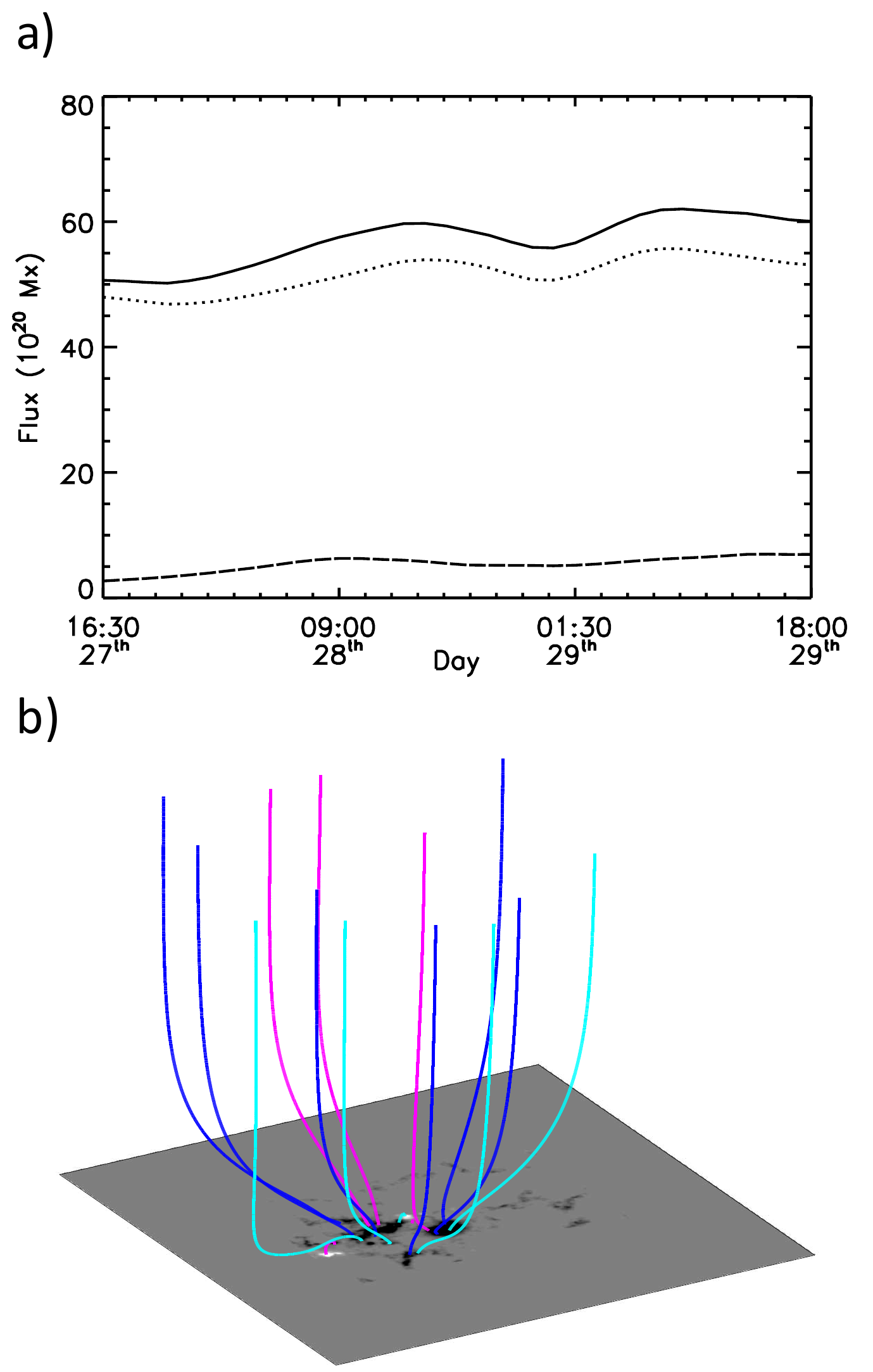}
\caption{(a) Graph of the variation of total flux (solid line), positive flux (dashed line) and absolute value of negative flux (dotted line) over the time period of the simulation. (b) Initial potential field configuration used in the simulation corresponding to a start time of 16:30:31\,UT on 27 March 2014.}
\label{fig:duncan1}
\end{figure}

The time period of the simulation ranges from 16:30:31\,UT on 27 March 2014 to 19:30:31\,UT on 29 March 2014 where the cadence of the magnetograms is taken to be 90 minutes. This time period is fully illustrated in Figure~\ref{fig:hmievo} where a selection of the magnetograms can be seen. From these panels it is clear that the area of interest is dominated by negative flux. While this is the case, from around 19:30:31\,UT on 27 March 2014 a new positive polarity of a bipole begins to appear and then grows strongly over the next two days. 
The variation of the total flux (solid line), unsigned negative flux (dotted line) and positive flux (dashed line) are seen in Figure~\ref{fig:duncan1} panel (a). From this plot it can be seen that both the positive and negative flux increase over time, characterised by the emergence of new flux. The numerical simulations of this active region are carried out in a computational box of $512^3$ grid points. Closed side boundary conditions are applied, but due to the dominance of negative flux open top boundary conditions are used. Use of the open top boundary condition means that no correction for flux balancing is required at the modelled photosphere. For use in the model, the time series of full disk magnetograms are de-rotated to disk center and a portion of size 300\,$\times$\,210 pixels is extracted and then centered at the lower boundary in the computational box. 
The normal field component on all computational grid points outside of the area on the magnetogram are set to zero. A potential magnetic field is then constructed from the initial magnetogram (Figure~\ref{fig:duncan1} panel (b)).  Due to the dominance of negative flux approximately 90$\%$ of the flux is open.

\begin{figure}[]
  \centering\includegraphics[scale=1.0,clip]{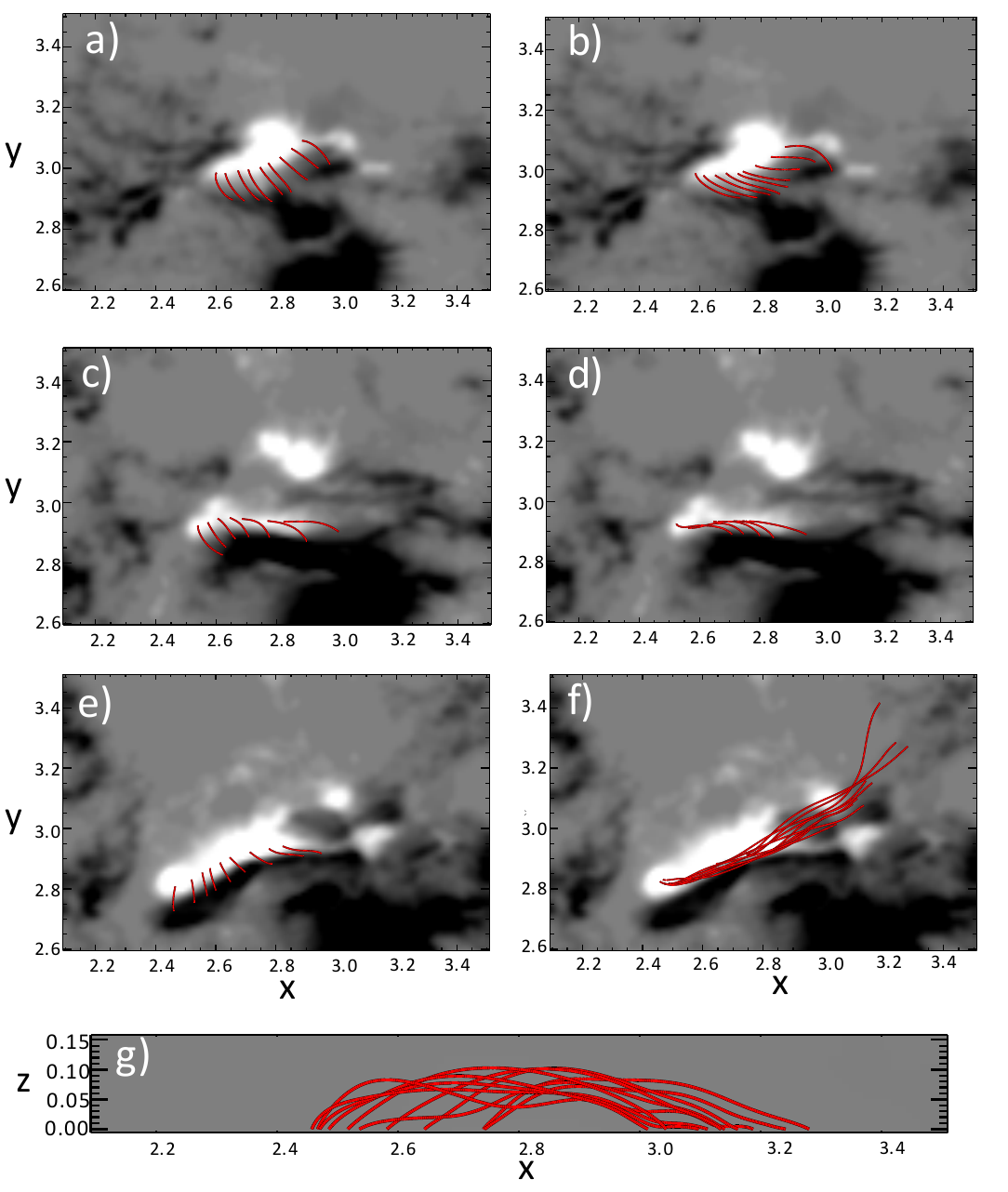}
 \caption{Connectivity of the field lines along the location of the filament at 09:30:31\,UT on 28 March 2014, (a) and (b), at 00:00:31\,UT on 29 March 2014, (c) and (d), and at 16:30:31\,UT on 29 March 2014, (e),(f) and (g). Panels (a) to (f) show the field lines from above superimposed on the magnetogram (white represents positive flux, black represents negative flux), while panel (g) shows the field lines from panel (f) viewed from the side. Finally panels (a), (c) and (d) do not include the additional helicity injection term, while panels (b), (d), (f) and (g) include additional helicity injection
at a rate of 3.75\,-\,5$\times10^{16}$\,Mx$^{2}$\,cm$^{-2}$\,s$^{-1}$ }
\label{fig:duncan2}
\end{figure}

In Figure~\ref{fig:duncan2} panel (a) the field lines produced in the quasi-static nonlinear force-free field simulation can be seen at 09:30:31\,UT on 28 March 2014. In this plot, white represents positive flux and black negative flux, where the field lines passing over the PIL are illustrated by the red lines. From this it is clear that these field lines do not exhibit a strong magnetic shear. Similar weak magnetic shear of the field is found in the corresponding field line connectivity at 00:00:31\,UT on 29 March 2014 (Figure~\ref{fig:duncan2} panel (c)) and 16:30:31\,UT on 29 March 2014 (Figure~\ref{fig:duncan2} panel (d)). Therefore none of these magnetic configurations produce a strongly sheared magnetic field along the PIL at the correct location or time that is representative of the field of a filament. Thus from this we find that for the present case the horizontal motions deduced from the magnetograms  do not inject enough non-potentiality or helicity into the coronal field to produce a strongly sheared magnetic field at the observed position of the filament. 

To investigate the possible origin of the strongly sheared non-potential magnetic structure of the filament a second series of simulations are carried out where an additional injection of helicity is included  at the photosphere. 
This is included through the addition of Equation (A1) from the paper of \cite{Mackay2014}. Inclusion of this term allows for the injection of either positive or negative helicity at the photosphere without changing the normal magnetic field. Figure~\ref{fig:duncan2} panel (b) shows the field lines at 09:30:31\,UT on 28 March 2014 when an average relative helicity density of $5 \times 10^{16}$\,Mx$^2$cm$^{-2}$s$^{-1}$ is injected into the corona throughout the simulation. It is clear that with this additional helicity injection the field lines now exhibit a strong shear that is directed to the left (sinistral) when standing on the positive polarity side of the PIL.  
 This strong shear along the PIL continues to build such that at 00:00:31\,UT on 29 March 2014 (Figure~\ref{fig:duncan2} panel d) there is an arcade lying directly along the PIL which then evolves into a magnetic flux rope by 16:30:31\,UT on 29 March 2014  (Figure~\ref{fig:duncan2} panel (f)). Figure~\ref{fig:duncan2} panel (g) shows a side view of the field lines in Figure~\ref{fig:duncan2} panel (f) where the final flux rope structure has approximately one turn along its length. Both the sign of helicity injection and chirality of the filament are of the dominant type found for the southern hemisphere.

To produce a magnetic structure that is consistent with the filament at the correct time and location an additional form of injection of magnetic helicity is required. Through carrying out a number of additional simulations with a variety of rates we find that the helicity density  injection rate has to be between $3.75-5 \times 10^{16}$\,Mx$^2$cm$^{-2}$s$^{-1}$ to reproduce the magnetic structure of the filament. 
It is important to point out that for the present simulations with open top boundary conditions, where initially over $90\%$ of the flux is open, the majority of this injected helicity is lost through the top boundary. The exact physical injection mechanism of this helicity at the present time is unknown. However since flux emergence is an important part of the 
evolution of the active region over the time period considered, one possible scenario for its origin is the transport of magnetic twist from the interior of the Sun to the atmosphere. While we only show results for a single box size, start time and initial condition, all three have been varied with and without additional helicity injection. In all cases similar results are found since the main magnetic feature of interest always emerges after the initial condition is constructed and as such is treated the same no matter the box size or initial condition used. While the present text gives a brief description of the results a future study will consider this in more detail along with a full description of the quantities calculated.

From this modelling we determine the behaviour of the magnetic field configuration in the active region leading up to the X-class flare. The presence and qualities of the magnetic flux rope revealed along the PIL, considered in conjunction with our observational results allow further conclusions to be drawn about the origin of the observed activity.

\section{Discussion}
The pre-flare period of the 29 March 2014 X-class flare shows very dynamic phenomena occurring in multiple layers of the solar atmosphere. This activity is observed in both spectroscopic and imaging data. 

Up to 40 minutes prior to flare onset, blue shifts are observed in multiple spectral emission lines by both IRIS and \textit{Hinode}/EIS in two of the three subregions (Regions B and C) of study chosen. The first of these areas, Region B, is located on the filament seen in AIA and IRIS SJI data. Region B is also the site of non-thermal line widths of 70\,km\,s$^{-1}$, observed in the coronal Fe {\sc{xii}} line. The plasma exhibiting this enhanced non-thermal velocity is also found to be blue-shifted. 

As detailed in Section 3.3 and shown in Figure~\ref{fig:regb}, within this region we observe strongly blue-shifted Si\,{\sc{iv}}, and He\,{\sc{ii}} emissions of up to 200\,km\,s$^{-1}$. These emissions are also highly transient, with velocities peaking at 16:54\,UT and 17:04\,UT whilst falling to lower values at 17:00\,UT. As earlier noted this presents an interesting offset between the times of peak blue shift occurring in the lower atmosphere and the corona where blue shifts are at a maximum at 17:00\,UT. This offset could be caused, for example, by the propagation of plasma through the atmosphere or be the result of two separate but related phenomena \textit{e.g.} small scale reconnection at different heights in the atmosphere. 

Region C also exhibits these strong transient blue shifts in Si {\sc{iv}} and He {\sc{ii}} data. As in the case of Region B, little activity is observed in these lines until $\approx$16:45\,UT. After this time 200\,km\,s$^{-1}$ blue shifts are observed until the onset of the X-flare. These blue shifts are very transient, particularly so in the Si\,{\sc{iv}} line profiles, where they ``switch on and off'' during the 40\,minutes prior to the flare at irregular intervals.  This activity suggests that the process causing these flows is not constant. In both these regions the chromospheric response as determined by the Mg\,{\sc{ii}} observations suggests that something is driving line broadening, particularly towards shorter wavelengths. However due to the complex nature of the chromospheric lines,  it is not a simple matter to say that these blue asymmetries are indicative of upflows. This effect is discussed in detail in \citet{kuridze2015}, who use simulations of H$\alpha$ emission to show that changes in the optical depth of the solar atmosphere driven by upflows can lead to absorption of red-wing photons at a higher altitude, producing a blue asymmetry. This effect has also been noted by \citet{kerr2016} to occur in Mg\,{\sc{ii}} lines.

In Section 3.4 we detail the locations of the observed plasma flows. In particular we highlight the IRIS Si\,{\sc{iv}} observations where the transient blue-shifted flows (Figure~\ref{fig:aiairis}) are found to coincide with an extended bright feature observed in AIA 193\,\AA\ data. These flows are fast with blue-shifted velocities up to a maximum of $\approx$200\,km\,s$^{-1}$, and with many regions exhibiting 100\,km\,s$^{-1}$ or upwards. These blue shifts and the related extended bright feature are very intriguing. The brightening and subsequent plasma velocities seen in both corona and transition region suggest that we are observing energy input into the region, possibly through reconnection. \citet{testa2014} undertook a study of small brightenings at the foot-points of active region loops using IRIS data to investigate non-thermal particle heating produced by nano-flares. In these bright points, they identified blue-shifted plasma with typical centroid velocities of $\approx$15\,km\,s$^{-1}$, and velocities of up to 40\,km\,s$^{-1}$ in the wings. These velocities are far slower than those that we have identified, leading us to conclude that nano-flare activity is likely not the driver of these flows and that we are not observing standard active region dynamics. 

The non-potential magnetic field modelling described in Section 3.5 shows that a magnetic flux rope is present in the location of the filament over an hour prior to the X-class flare. The presence of a magnetic flux rope is also supported by \citet{yang2016}, whose modelling identified a flux rope in the active region from 09:00\,UT. The inferred presence of a flux rope within the active region before the flare allows us to confine our ideas of what could be driving the observed flows. Strongly blue-shifted flows of up to 200\,km\,s$^{-1}$ in the transition region have been observed in pre-flare IRIS data by \citet{cheng2015}. In this paper the authors presened observations of two active regions prior to flare events. Within these areas studied, blue shifts of 200\,km\,s$^{-1}$ were identified to  occur in the presence of filaments. The authors interpreted these fast up-flows to be the result of magnetic reconnection occurring between two smaller flux ropes, to create a larger flux rope. The authors also identified down-flows at the foot points of the final flux rope. These strong flows were observed in Ca\,{\sc{ii}}, Mg\,{{\sc{ii}}} and Si\,{\sc{iv}}. \citet{cheng2015} did not investigate the coronal response to this activity, but comparing this work to our own observations, we note that the presence of strong flows located in the centre of the filament/flux rope, and observed from the chromosphere through to the corona, compares well. We do not observe the foot-points of the filament/flux rope with the spectrometers and so cannot comment on whether this piece of the \cite{cheng2015} interpretation is present in our results. The presence of the flux rope in the region of study is another piece of evidence in favour of using the \cite{cheng2015} model to interpret this pre-flare behaviour.
 \begin{figure} 
 \centerline{\includegraphics[width=0.7\textwidth,clip=]{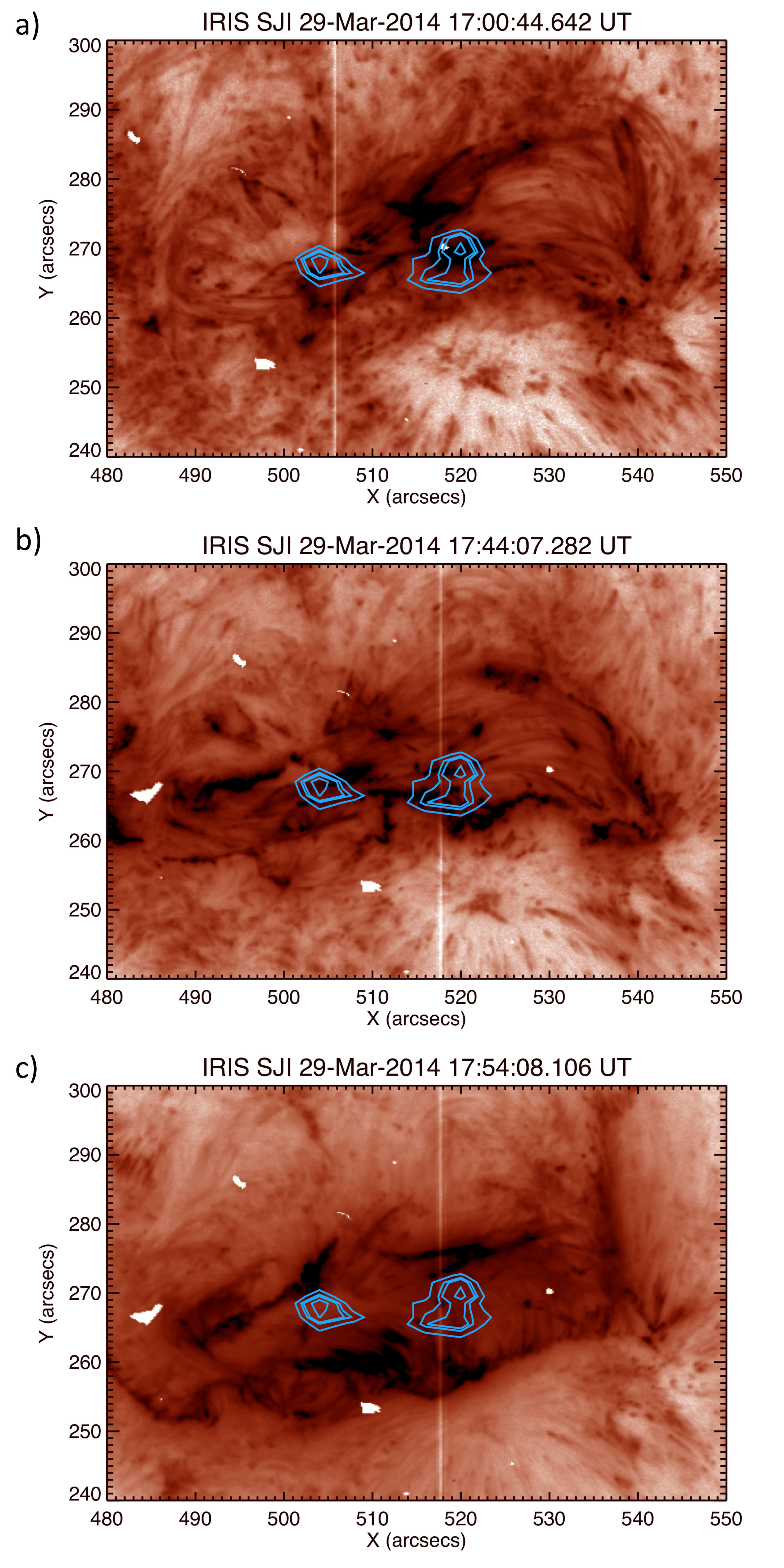}}
 \caption{Comparison of the position of blue shifts observed in \textit{Hinode}/EIS Fe\,{\sc{xii}} data at $-100$\,km\,s$^{-1}$ with flare ribbons identified in IRIS slit-jaw imager 1400\,\AA\ channel (an inverted colour table has been used). Panel (a) shows the position of the position of the blue shifts at the time they occurred. These areas of blue shift are aligned with brightenings also visible in AIA 193\,\AA\ data. Panel (b) shows the blue shifts superimposed onto SJI data during flaring. This image is chosen as it is the closest image to the flare peak that is not saturated. Flare ribbons are visible and appear to lie close to the brightenings seen in earlier SJI and AIA 193\,\AA\ data. Panel (c) shows the situation post flare. We can clearly see the flare ribbons have expanded outwards from their initial positions.}
 \label{fig: ribbons}
 \end{figure}
 \begin{figure}[]
\centering
\includegraphics[width=1.0\textwidth]{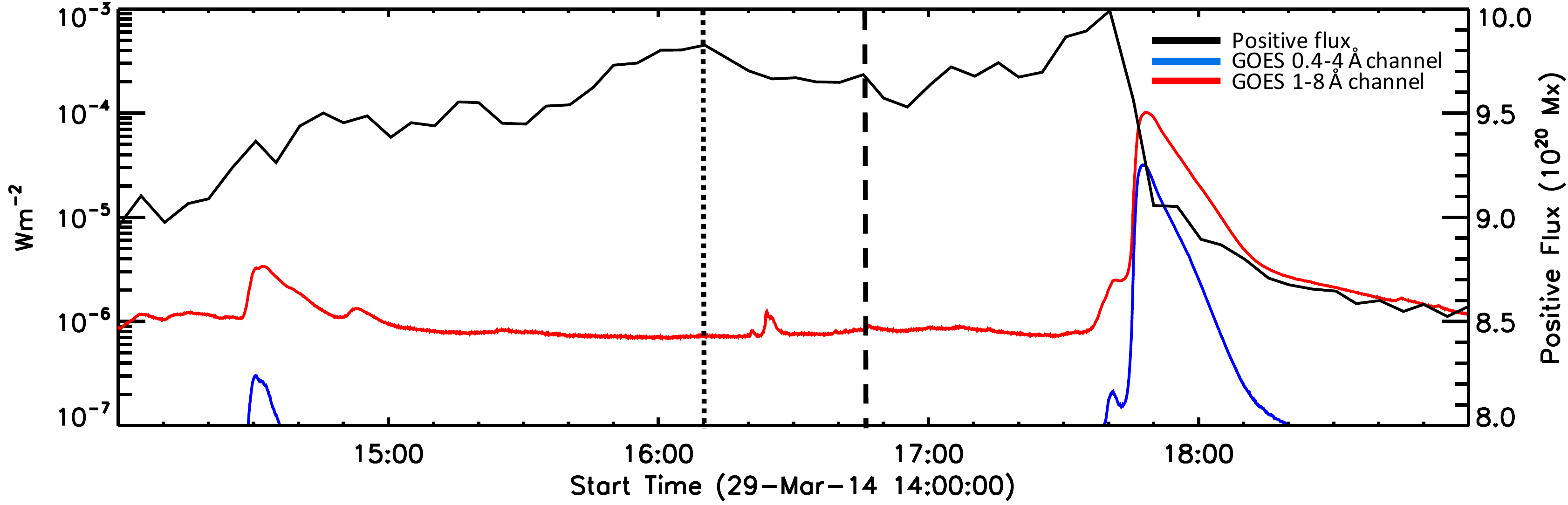}
\caption{Evolution of positive flux with time over-plotted on the GOES light curve. We can see that over the period of joint EIS/IRIS observation there is sustained flux emergence into the region of study. This emergence continues until 16:10\,UT, highlighted by the dotted line. The nadir of this drop is coincident with the C-class flare at 16:24\,UT. The flux then stays constant until 16:46\,UT, where a second smaller drop in flux is observed (dashed line). This drop in magnetic flux is broadly coincident with the onset of the flows described in Section 3.2. This decrease in flux recovers quickly to its previous level. Again the flux remains roughly constant until 17:30\,UT where a large increase in the flux is observed prior to the steep drop coincident with the X-flare at 17:35\,UT. }
\label{fig: fluxem}
\end{figure}

Kink instability is another possible mechanism through which solar flares can be triggered. If the kink instability were to be the driver for the flare we observe we would expect to identify both red and blue-shifted plasma along the flux rope \citep[\textit{e.g.},][]{williams2009}. From our spectroscopic observations we identify only blue shifts along the length of the filament. This casts significant doubt on the kink instability being responsible for the activity. The weakly twisted nature of the magnetic flux rope identified through our modelling also serves to further rule out the kink instability as the driver. We determine that the level of twist in the flux rope is insufficient to become kink unstable. We therefore conclude that thekink instability is not the driver of the observed activity due to the findings of our spectral observations and modelling.
 
 Tether-cutting reconnection and magnetic breakout may also be considered as possible explanations for the strong flows and extended pre-flare brightenings observed. Activity associated with the breakout model would be expected to occur away from the filament. As the observed activity is seen to lie along the filament we conclude that it is unlikely that reconnection due to magnetic breakout is the driver of this activity within the spectrometer fields of view. Tether-cutting reconnection proves to be a better candidate to explain this activity as related brightenings have been observed close to filaments \citep{Warren2001}. The work of \citet{Kleint2015} showed the filament in this active region to be slowly rising in the hours before the eventual eruption, which would also be consistent with the tether cutting model.

As the extended bright features occur in the region of the flare and filament eruption, we investigate whether there could be a relation to the flare ribbons. In Figure~\ref{fig: ribbons} we show \textit{Hinode}/EIS Fe\,{\sc{xii}} data at $-100$\,km\,s$^{-1}$, observed at 17:00\,UT and overlaid on to IRIS slit jaw images in the 1400\,\AA\ pass band at three separate times. In panel (a) we see the extended brightenings at 17:00\,UT, that are discussed in Figure~\ref{fig:aiairis}, linking the two regions of \textit{Hinode}/EIS blueshift. During the flare at 17:44\,UT (panel (b)) we can see that the appearance of flare ribbons and their locations. There appear to be ribbon brightenings in close proximity to the locations of the \textit{Hinode}/EIS blue-shifted regions. Strong brightenings are also clearly seen to lie along the path of the filament as it appeared prior to its eruption. The slit jaw imager data saturates as the peak of the flare is reached. Panel (c) shows the situation at 17:54\,UT, after saturation of the data has passed. Here we see that the flare ribbons are more obvious and have expanded outwards from their earlier positions. Flare ribbons are commonly observed during a flare, but have been identified prior to the impulsive phase. \citet{fletcher2013} identified flare ribbons appearing to lie within a filament minutes prior to flaring after which they underwent outward expansion. The locations of the extended brightenings in panel (a) and the flare ribbons in panel (b) are both very close to the location of the filament. This could suggest that the extended brightening may have a relation to the flare ribbons, however due to their appearance $\approx$40 minutes before onset of flaring, it is highly doubtful that they are actually ribbons themselves.

The work of \citet{kusano2012} and \citet{bamba2013} proposed and provided observational evidence for small scale flux emergence at the PIL of an active region being related to the triggering of flares. For the active region studied in this paper, Figure~\ref{fig: fluxem} shows the evolution of positive magnetic flux compared to GOES soft X-ray flux between 14:00\,UT and 19:00\,UT on 29 March 2014. We see a general trend of increasing magnetic flux over time. A gradual decrease in magnetic flux is observed from 16:10\,UT, marked by the dotted line in Figure \ref{fig: fluxem}. The nadir of this decrease is coincident with the 16:24\,UT C-flare. The flux level then plateaus for $\approx$25 minutes until another drop is observed (marked by the dashed line in Figure \ref{fig: fluxem}). This second decrease coincides with the onset of the strong flows that we have observed throughout the solar atmosphere around 17:00\,UT. The flux level increases once more to a maximum value of $9.99\times10^{20}$\,Mx just prior to the onset of the X-flare at 17:35\,UT. At this time a rapid decrease in magnetic flux is then observed in conjunction with the impulsive rise in GOES soft X-ray flux. This behaviour is consistent with the events designated as opposite polarity (OP) in \citet{bamba2013}, where small bipole fluxes that emerge along the PIL are of opposite polarity to the active region into which they emerge. In this scenario, magnetic flux must rise to a critical level and rapidly decrease when the flare is triggered by an instability in the system. 
The apparent decrease in magnetic flux at the time of the strong flows also indicates that there is some process occurring in the same region that is leading to the loss of flux. 
This could be due to flux cancellation, with reconnection driving the observed flows in the upper atmosphere and the subsequent submergence of the resulting small post reconnection loops being the observed flux cancellation. Alternatively, the decrease in magnetic flux at the time of the observed pre-flare activity could be due to plasma flows in the atmosphere, or a heating process exerting a pressure upon the magnetic field. This could the change the inclination of the magnetic field resulting in the observed decrease in magnetic flux due to line-of-sight effects.
 
\section{Conclusions}
In this work we carry out the first simultaneous spectroscopic study of pre-flare activity in the solar atmosphere from the corona to the chromosphere. Using observations from \textit{Hinode}/EIS and IRIS we have identify the presence of strongly blue-shifted plasma flows with velocities of up to 200\,km\,s$^{-1}$ in Si\,{\sc{iv}} emissions from the transition region. These strongly blue-shifted flows also appear to be related to an extended bright feature observed to lie along the filament by SDO/AIA and IRIS SJI data. In our efforts to investigate these observed  features, we undertake non-potential magnetic field modelling of the region detailed in Section 3.5. This reveals the presence of a weakly twisted flux rope over one hour prior to the flare and filament eruption. Section 4 details how these results relate to existing flare trigger models. The models which do not fit with our observations are as follows. i) Kink Instability: due to the weakly twisted nature of the flux rope and the absence of evidence of untwisting\,(red and blue shifts along the filament) we determine that it is highly unlikely that the activity is driven by the kink instability. ii) Breakout reconnection: within the spectrometer fields of view, all brightenings and flows are observed to occur close to or within the filament. This therefore rules out the likelihood of this activity being explained by breakout reconnection within the field of view available. However, as the field of view of both spectrometers is small, we cannot rule out breakout reconnection occurring remotely from the area we observe. 

More probable explanations for these flows and brightenings are that they are related to the following phenomena. i) Reconnection in the flux rope: the strongly blue-shifted plasma that we are observing in the centre of the flux rope maybe driven by a process akin to that described by \citet{cheng2015}. ii) Early onset of flare reconnection: the positions of the plasma flows and brightenings in relation to the flare ribbons are investigated. It is found that the positions of the brightenings can be related to the flare ribbons that appear during the flare. This would imply that we are observing the onset of flare reconnection 40 minutes prior to the main flare onset. iii) Tether-cutting reconnection: the position of the strong flows and brightenings along the filament suggests that the observed activity may be the result of tether-cutting reconnection.

The pre-flare environment is very complex and not all activity related to the different trigger mechanisms is unique. The strong flows that we observe could be indicative of several possible flare drivers. We are therefore unable to provide conclusive evidence of one mechanism that drives the onset for the flare. Due to the complex nature of the solar atmosphere at the time the activity is observed it is perhaps likely that some combination of all the possible trigger mechanisms drives the observed activity. 
%

%
 \begin{acks}
 IRIS is a NASA Small Explorer mission developed and operated by LMSAL with mission operations executed at NASA Ames Research center and major contributions to downlink communications funded by ESA and the Norwegian Space Centre.
\textit{Hinode} is a Japanese mission developed and launched by ISAS/JAXA, collaborating with NAOJ as a domestic partner, and NASA and STFC\,(UK) as international partners. Scientific operation of the \textit{Hinode} mission is conducted by the \textit{Hinode} science team organised at ISAS/JAXA.
MMW and SD acknowledge STFC for support via their PhD Studentships.
DML is an Early-Career Fellow, funded by the Leverhulme Trust.

 \end{acks}

\hfil \\
\noindent
\textbf{Disclosure of Potential Conflicts of Interest} The authors declare that they have no conflicts of interest.

%
%
 \bibliographystyle{spr-mp-sola}
 \bibliography{references}  
 
%
%
%
%

\end{article} 
\end{document}